\documentclass[utf8]{FrontiersinHarvard} 

\usepackage{url,hyperref,lineno,microtype,subcaption}
\usepackage[onehalfspacing]{setspace}
\usepackage{siunitx}
\usepackage{multirow}
\usepackage[flushleft]{threeparttable}

\def\keyFont{\fontsize{8}{11}\helveticabold }
\def\firstAuthorLast{Fang} 
\def\Authors{Kun Fang\,$^{1,*}$}
\def\Address{$^{1}$Key Laboratory of Particle Astrophysics, Institute of High Energy Physics, Chinese Academy of Sciences, Beijing, People’s Republic of China}

\def\corrAuthor{Kun Fang}
\def\corrEmail{fangkun@ihep.ac.cn}

\begin{document}
\onecolumn
\firstpage{1}

\title[Galactic Pulsar Halos]{Gamma-ray Pulsar Halos in the Galaxy} 

\author[\firstAuthorLast ]{\Authors} 
\address{} 
\correspondance{} 

\extraAuth{}

\maketitle

\begin{abstract}


Pulsar halos are extended gamma-ray structures generated by electrons and positrons escaping from pulsar wind nebulae (PWNe), considered a new class of gamma-ray sources. They are ideal indicators for cosmic-ray propagation in localized regions of the Galaxy and particle escape process from PWNe. The cosmic-ray diffusion coefficient inferred from pulsar halos is more than two orders of magnitude smaller than the average value in the Galaxy, which has been arousing extensive discussion. We review the recent advances in the study of pulsar halos, including the characteristics of this class of sources, the known pulsar halos, the possible mechanisms of the extremely slow diffusion, the critical roles of pulsar halos in the studies of cosmic-ray propagation and electron injection from PWNe, and the implications on the problems of the cosmic positron excess and the diffuse TeV gamma-ray excess. Finally, we give prospects for the study in this direction based on the expectation of a larger sample of pulsar halos and deeper observations for bright sources.

\tiny
 \keyFont{ \section{Keywords:} cosmic rays, cosmic ray propagation,  supernova remnants, pulsar, pulsar wind nebulae} 
\end{abstract}

\section{Introduction}
In 2007, the Milagro Gamma-ray Observatory reported an extended TeV gamma-ray source around the Geminga pulsar \citep{Abdo:2007ad}. At that time, the source was considered a TeV pulsar wind nebula (PWN) associated with Geminga \citep{Yuksel:2008rf}. However, the $\sim\ang{3}$ extension of the source is hard to understand, which is dozens of times larger than the bow-shock PWN of Geminga observed at X-ray wavelengths \citep{2003Sci...301.1345C}. A decade later, the High-Altitude Water Cherenkov Observatory (HAWC) updated the measurement of this gamma-ray source with better sensitivity and angular resolution \citep{Abeysekara:2017old}. The gamma-ray morphology shows radial symmetry and can be well interpreted by the diffusion model, indicating that the source is more likely to be generated by electrons and positrons\footnote{\textit{Electrons} will denote both electrons and positrons hereafter if not specified.} that have escaped from the central PWN and are freely diffusing in the interstellar medium (ISM) \citep{Hooper:2017gtd,Fang:2018qco,Profumo:2018fmz,Tang:2018wyr,Giacinti:2019nbu}. This new source type is known as the gamma-ray pulsar halo, TeV halo, or ICS halo. We will use the name pulsar halo throughout this paper.

Pulsar halos are ideal probes for cosmic-ray (CR) propagation in localized regions of the Galaxy. High-energy electrons released by the PWNe inverse Compton (IC) scatter homogeneous background photons to produce the gamma-ray halos, so the gamma-ray morphologies unambiguously trace the propagation of the parent electrons. The most intriguing result is that the inferred electron diffusion coefficient is several hundred times smaller than the average CR diffusion coefficient in the Galaxy \citep{Abeysekara:2017old,Aharonian:2021jtz}. It complicates the image of Galactic CR propagation and has significant impacts on some key issues of CRs, such as the origin of the positron excess \citep{PAMELA:2008gwm,Fermi-LAT:2011baq,AMS:2013fma} and the diffuse TeV gamma-ray excess \citep{Milagro:2005xqq,Abdo:2008if,TibetASgamma:2021tpz}.

Pulsar halos are essential for the study of electron injection from PWNe \citep{Fang:2022mdg,2022arXiv220713533F}. As the gamma-ray halos are generated by the escaping electrons from PWNe, the energy spectrum of the electrons injected from PWNe into the ISM (injection spectrum for short) should be inferred from the halo spectrum rather than the spectrum of PWNe themselves. The injection spectrum could be further used to estimate the acceleration limit of PWNe \citep{2019BAAS...51c.311F}. Unlike the high-energy spectral cutoff of X-ray PWNe that is determined by both the maximum electron energy and the magnetic field of PWNe, the gamma-ray spectrum of pulsar halos can unambiguously indicate the electron spectral cutoff. Pulsar halos are also proposed to search for invisible pulsars \citep{Linden:2017vvb}.

This work reviews the recent advances in the study of pulsar halos. In Section~\ref{sec:character}, we introduce the characteristics of this class of sources and the known pulsar halos. In Section~\ref{sec:origin}, current theories about the origin of pulsar halos are reviewed. We present the significance of pulsar halos to the studies of CR propagation and electron injection from PWNe in Sections~\ref{sec:prop} and \ref{sec:injection}, respectively, before presenting the implications for interpreting the positron excess in Section~\ref{sec:positron} and the diffuse TeV gamma-ray excess in Section~\ref{sec:diffuse}. We look forward to the future studies of pulsar halos in Section~\ref{sec:perspect} and give the summary points in Section~\ref{sec:sum}. One may also refer to other recent reviews of pulsar halos \citep{Lopez-Coto:2022igd,2022arXiv220704011L}, which have different focuses from the present work.

\section{Characteristics of Pulsar halos}
\label{sec:character}

\subsection{Criteria of a Pulsar Halo}
\label{subsec:CTs}
We first present several criteria for a pulsar halo and then introduce the characteristics of pulsar halos by explaining these criteria. Some of them have been proposed in \citet{2022arXiv220713533F}.

\begin{enumerate}
    \item The TeV pulsar halo of a visible pulsar should have spatial coincidence with that pulsar.
    \item The spin-down luminosity of the central pulsar should be large enough to generate the pulsar halo. 
    \item The extension of a gamma-ray pulsar halo should be significantly larger than that of the X-ray PWN if the X-ray PWN is observable. 
    \item The morphology of a pulsar halo should be interpreted well by the diffusion-loss propagation of electrons with a reasonable diffusion coefficient.
\end{enumerate}

After escaping from the PWN, the electron propagation in ISM is generally described by the diffusion process, so the gamma-ray halo should be centered at the host pulsar, which is a more specific description of criterion 1 (CT~1). Even if the diffusion environment is asymmetric, the flux peak of the gamma-ray halo should be coincident with the pulsar. CT~1 holds for $E_\gamma\gtrsim1$~TeV \citep{DiMauro:2019hwn}, where $E_\gamma$ is the gamma-ray energy. For example, HESS~J1026$-$582 is a TeV gamma-ray source once considered a possible pulsar halo \citep{DiMauro:2019hwn}. However, the offset between the source centroid and the assumed host pulsar PSR~J1028$-$5819 is significantly larger than the Gaussian extension of the source \citep{HESS:2010aa}, which cannot pass CT~1. On the other hand, a GeV pulsar halo may deviate from the position of the associated pulsar owing to the pulsar's proper motion \citep{Johannesson:2019jlk,DiMauro:2019yvh,ZhangZhangYi:2021kzq}. 

CT~2 is essential to judge the association between a gamma-ray source and a nearby pulsar. For example, HESS~J1632$-$478 was once connected with a nearby middle-aged pulsar PSR~J1632$-$4757 \citep{DiMauro:2019hwn}. However, the gamma-ray luminosity of HESS~J1632$-$478 is significantly larger than the spin-down energy of the pulsar, and this source is more likely to be a PWN associated with a much younger pulsar \citep{Balbo:2010pf}. CT~2 is also helpful for constraining theoretical models of pulsar halos. A model will be disfavored if the required injection energy is larger than the pulsar can provide. An example is shown in Section~\ref{subsec:relativistic}. 

According to the standard picture of PWN evolution \citep{Gaensler:2006ua}, a pulsar and its original PWN are located at the center of the parent supernova remnant (SNR) in the early age. The expanding PWN encounters the SNR reverse shock and is disrupted after a time of $\sim7$~kyr, leaving a relic PWN. Meanwhile, the pulsar leaves the birthplace and then the relic PWN owing to the kick velocity (typically $400-500$~km~s$^{-1}$) and finally escapes into the ISM after a time of $\sim$50~kyr. Only then can the gamma-ray halo morphology indicates the CR propagation in the ISM. Thus, we may give a preference to the pulsars older than $\sim$50~kyr for the study of pulsar halos. For a gamma-ray source associated with a much younger pulsar, we should be more cautious as it is likely to be a pure PWN or a source in the mixed state of a pulsar halo and a relic PWN. Older pulsars with $\tau\gtrsim1$~Myr may not hold an observable pulsar halo owing to the low spin-down energy, so pulsar halos are generally considered to be generated by middle-aged pulsars with $\tau\sim100$~kyr. However, a recycled millisecond pulsar with a very large characteristic age ($\tau\gtrsim50$~Myr) can be as bright as middle-aged pulsars and may also hold a pulsar halo \citep{Hooper:2018fih,Hooper:2021kyp}.

When a pulsar travels outside its SNR (or in the outer part of its SNR), the pulsar motion can be supersonic and drive a bow-shock PWN that is significantly different from its original PWN. The size of the bow-shock PWN is limited by the ram pressure caused by the pulsar motion, which is $\lesssim1$~pc and not expected to expand with time \citep{Gaensler:2006ua}. The bow-shock PWN in X-ray and the corresponding TeV halo are generated by electrons with similar energies. Thus, the X-ray PWN can be regarded as the source region of the TeV halo and should then be significantly smaller than the TeV halo, resulting in CT~3. On the other hand, a TeV source that has a comparable size to the X-ray PWN is more likely a gamma-ray PWN \citep{HESS:2012quc}. CT~3 is important in distinguishing a pulsar halo from a PWN.

It is also necessary to distinguish between a pulsar halo and a gamma-ray source with a hadronic origin. A nearby hadronic accelerator and a reasonable gas distribution in the source region are the essential conditions of a hadronic origin. Moreover, these two scenarios predict different gamma-ray morphologies \citep{Yang:2021iwe}. The morphology of a pulsar halo should be interpreted well by the diffusion-loss propagation of electrons, which is CT~4. The derived diffusion coefficient may also help to constrain the origin of the source. Bohm limit is considered the lower limit of the diffusion coefficient and is written as $1/3r_gc$, where $r_g$ is the gyro-radius of CR particles. If the derived diffusion coefficient is smaller than the Bohm limit under the typical magnetic field strength in the ISM, the gamma-ray structure may not be a pulsar halo, or the pulsar distance is significantly underestimated. Besides, CT~4 could be valuable in finding pulsar halos of invisible pulsars. 

The standard diffusion scenario fits well with the current morphology measurements of pulsar halos, and CT~4 may play a significant role in identifying pulsar halos at present. However, we cannot rule out the possibility that more sophisticated propagation models are needed to interpret more accurate observations in the future.

\subsection{Observed Pulsar Halos}
\label{subsec:halos}
\begin{figure}[t!]
\includegraphics[height=4.7cm,width=5.1cm]{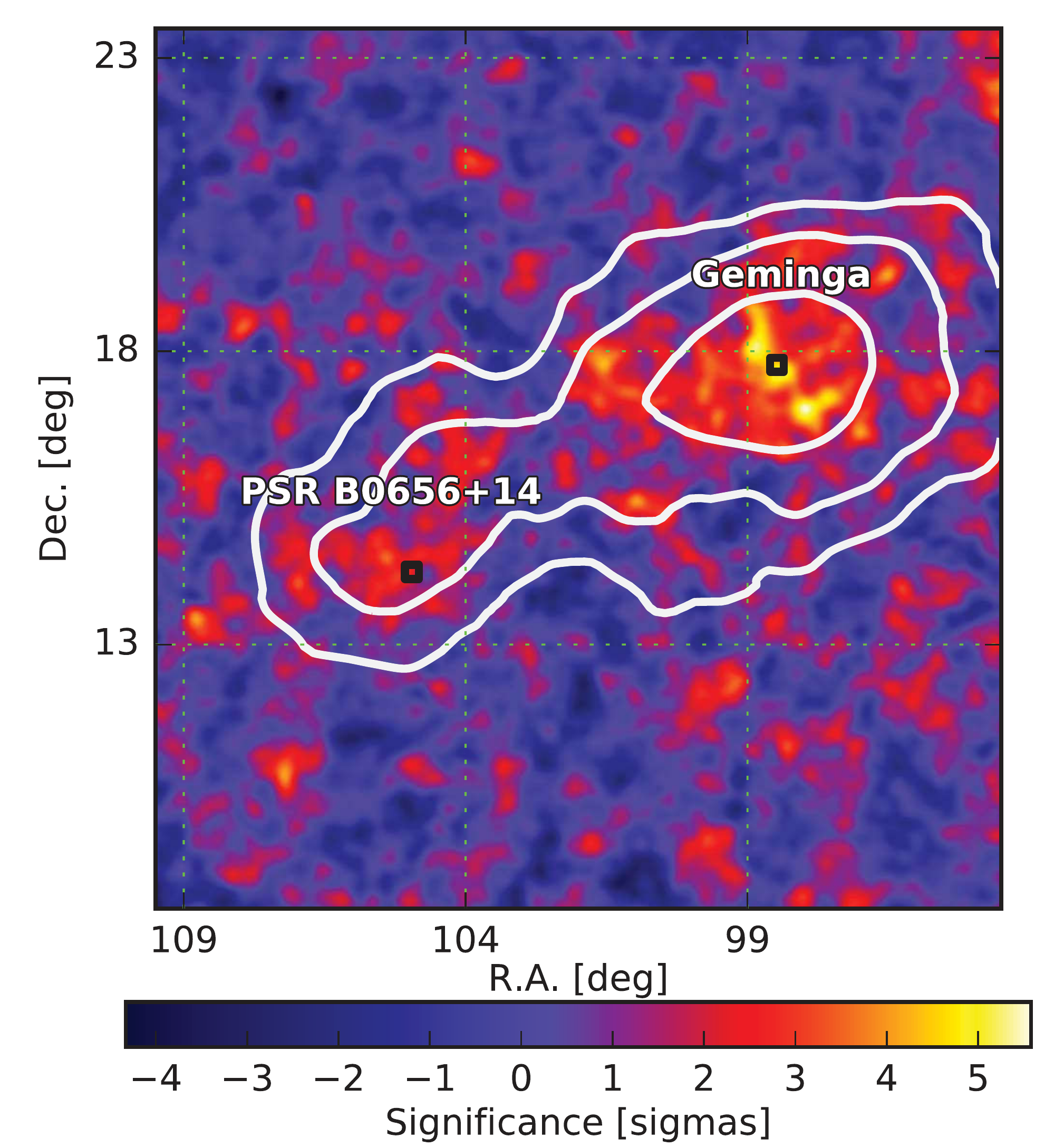}
\includegraphics[width=5.3cm]{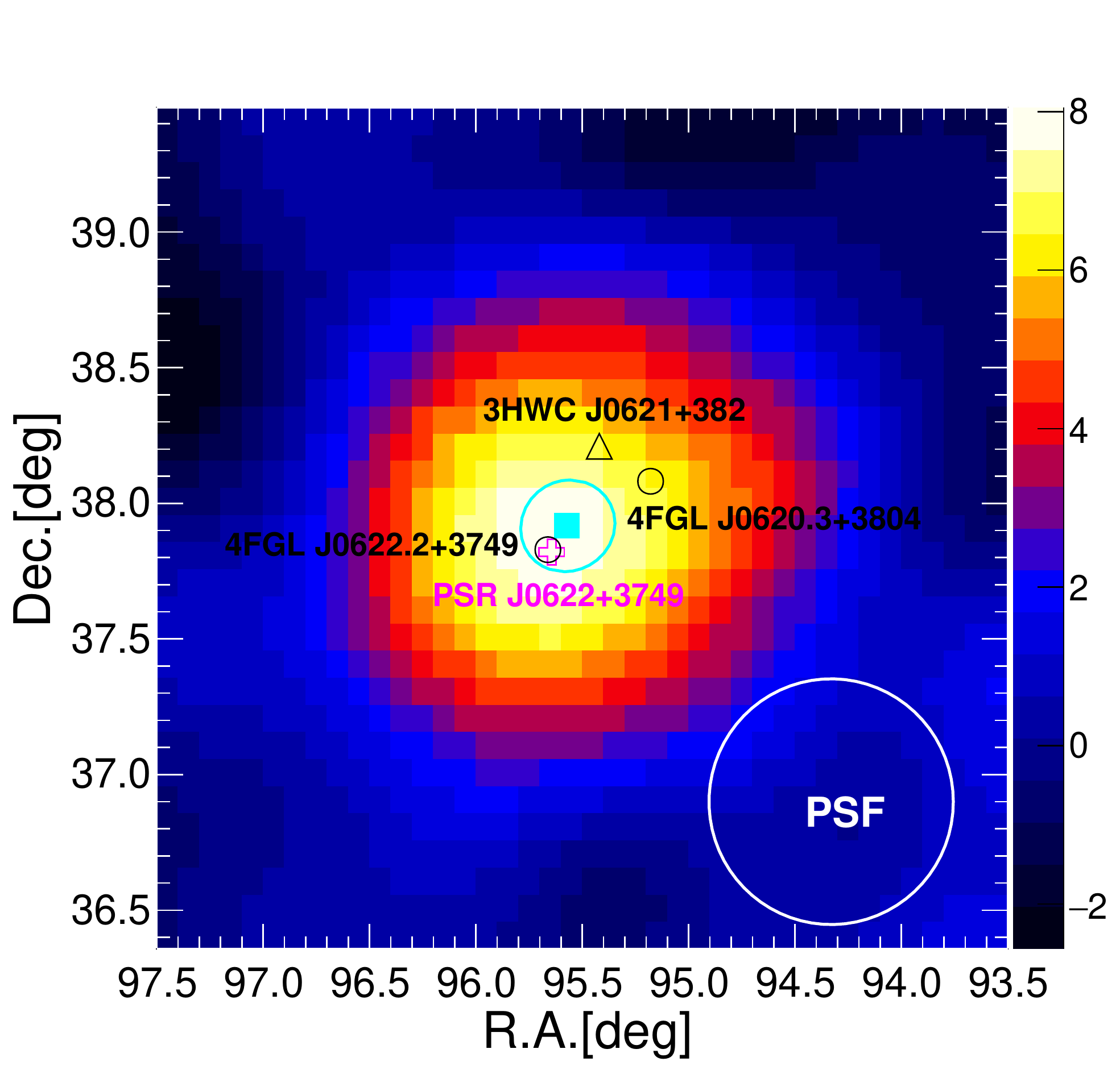}
\includegraphics[width=5.7cm]{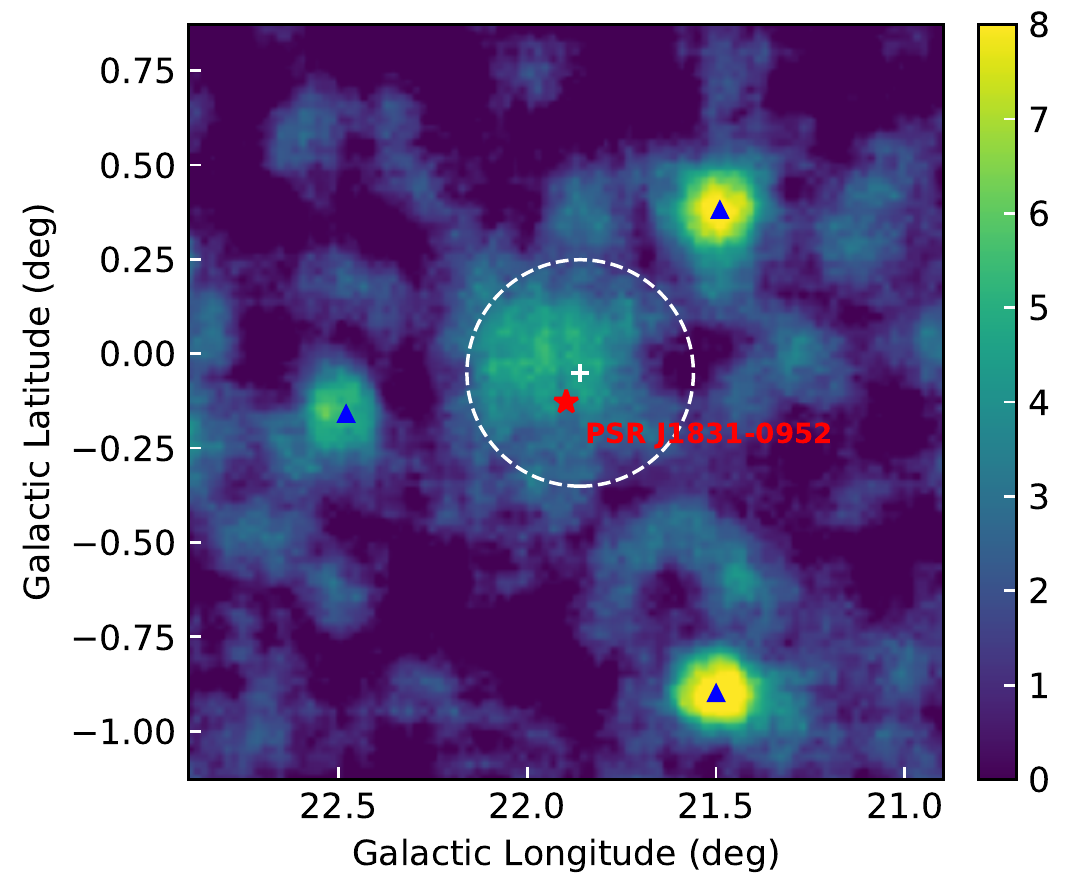}
\caption{Significance maps of the known gamma-ray pulsar halos. \textbf{Left:} Geminga and Monogem halos observed by HAWC, taken from \citet{Abeysekara:2017old} with permission from Science. \copyright 2017 AAAS. \textbf{Middle:} LHAASO~J0621$+$3755 observed by LHAASO-KM2A, taken from \citet{Aharonian:2021jtz} with permission from Physical Review Letters. \copyright 2021 APS. \textbf{Right:} Significance map of HESS~J1831$-$098 given by the H.E.S.S. Galactic plane survey \citep{HESS:2018pbp} with the integration radius of \ang{0.1}, taken from \citet{2022arXiv220713533F}. One may refer to the original papers for the meaning of the marks in the figures.}
\label{fig:halos}
\end{figure}

At present, about ten gamma-ray sources are proposed to be pulsar halos or candidates for pulsar halos \citep{Albert:2020fua,Aharonian:2021jtz}. Four of these sources have been well studied as pulsar halos: the Geminga halo, the Monogem halo, LHAASO~J0621$+$3755, and HESS~J1831$-$0952. Figure \ref{fig:halos} summarizes the significance maps of these sources. Information on the associated pulsars and two important parameters derived from the observations of the pulsar halos are summarized in Table \ref{tab:para}. The Geminga halo is the brightest pulsar halo and is regarded as the canonical source of this class. Geminga is one of the brightest gamma-ray pulsars \citep{Fermi-LAT:2013svs} and is close to Earth. It has received much attention from the CR community as it is a likely source of the cosmic positron excess. We continue this topic in Section~\ref{sec:positron}. 

\renewcommand{\arraystretch}{1.5}
\begin{table}[t!]
 \centering
 \caption{Pulsar parameters of known pulsar halos and parameters inferred from pulsar halos.}\label{tab:para}
 \vspace{3mm}
 \begin{tabular}{cccccccc}
  \hline
  \hline
  \multirow{3}{*}{Name} & \multicolumn{5}{c}{Parameters of associated pulsars} & \multicolumn{2}{c}{Derived parameters} \\
  \cline{2-8}
  {} & Gl & Gb & $T$ & $D$ & $L$ & $D_{100}$ & $\eta$ \\
  {} & [deg] & [deg] & [kyr] & [kpc] & $10^{34}$ [erg s$^{-1}$] & [cm$^2$ s$^{-1}$] & [\%] \\
  \hline
  Geminga halo & $195.13$ & $4.27$ & $342$ & $0.25^a$ & $3.2$ & $4.6\times10^{27}$~$^c$ & $5^c$ \\
  Monogem halo & $201.11$ & $8.26$ & $111$ & $0.288$ & $3.8$ & $1.5\times10^{28}$~$^d$ & $4^d$ \\
  LHAASO J0621$+$3755 & $175.88$ & $10.96$ & $208$ & $1.6^b$ & $2.7$ & $2.3\times10^{27}$~$^c$ & $40^c$ \\
  HESS J1831$-$098 & $21.90$ & $-0.13$ & $128$ & $3.68$ & $110$ & $9.0\times10^{27}$~$^e$ & $7^e$ \\
  \hline
 \end{tabular}
 \begin{tablenotes}
  \small
  \item \textbf{Notes.} Columns 2-6 are the Galactic longitude, Galactic latitude, characteristic age, distance, and spin-down luminosity, respectively. Pulsar parameters are taken from the ATNF pulsar catalog \citep{Manchester:2004bp} unless specified. Column 7 is the diffusion coefficient normalized at 100~TeV. Column 8 is the needed conversion efficiency from the pulsar spin-down energy to the electron energy of pulsar halo. $\eta$ is obtained by assuming an injection spectrum with a ECPL form, except the case of Monogem, where $\eta$ is obtained by assuming a simple power-law injection. $\eta$ will be significantly smaller if the ECPL injection is adopted for Monogem.
  \item References: $^a$\citet{2007Ap&SS.308..225F}; $^b$\citet{Parkinson:2010xf}; $^c$\citet{Bao:2021hey}; $^d$\citet{Abeysekara:2017old}; $^e$\citet{2022arXiv220713533F}.
 \end{tablenotes}
\end{table}

The Monogem halo is associated with PSR~B0656$+$14, known as the Monogem pulsar. The Monogem halo was reported by HAWC together with the Geminga halo \citep{Abeysekara:2017old} and is also a nearby source. However, it is significantly dimmer than the Geminga halo, so the diffusion coefficient in the halo is not accurately constrained. 

LHAASO~J0621$+$3755 is the pulsar halo of PSR~J0622$+$3749 and is the first pulsar halo reported by the Large High-Altitude Air Shower Observatory (LHAASO) \citep{Aharonian:2021jtz}. Intriguingly, this source could be far away from the Galactic plane ($\approx300$~pc above the plane), which is distinct from the other known pulsar halos. 

HESS~J1831$-$098 is the pulsar halo of PSR~J1831$-$0952. This source was first discovered by the High Energy Spectroscopic System (H.E.S.S) and was considered an old PWN at that time \citep{Sheidaei:2011vg}. However, \citet{2022arXiv220713533F} point out that it is more reasonable to regard it as a pulsar halo as it passes all the criteria introduced above. Although this source is more than $10$ times farther than Geminga, the significantly stronger injection power ensures its visibility. 

The four observed pulsar halos all meet CT~1-4. The separations between the centroids of the gamma-ray halos and the pulsars are all significantly smaller than the extensions of the corresponding halos \citep{Albert:2020fua,Aharonian:2021jtz,2022arXiv220713533F}, indicating good spatial coincidence. Criteria about conversion efficiency and pulsar age are all satisfied, as shown in Table~\ref{tab:para}. Geminga, Monogem, and PSR~J1831$-$0952 all hold X-ray PWNe, which are at least dozens of times smaller than the corresponding TeV halos \citep{Posselt:2016lot,Birzan:2015puz,2019ATel12463....1A}. No clear x-ray PWN is found for PSR~J0622$+$3749, possibly owing to the faintness of the PWN \citep{Aharonian:2021jtz}. Finally, all the TeV pulsar halos can be interpreted with the standard diffusion framework under the current measurement accuracy. The measured diffusion coefficients are summarized in Table~\ref{tab:para}.

For other possible pulsar halos, such as HAWC~J0543+233 \citep{2017ATel10941....1R} and HAWC~J0635+070 \citep{2018ATel12013....1B}, the information reported is still not comprehensive enough, and we cannot test these candidates with all our criteria at present.

\section{Possible Origins of Pulsar halos}
\label{sec:origin}
A pulsar halo is not expected under the average CR diffusion coefficient in the Galaxy as the electrons escaping from the PWN will spread out rapidly. \citet{Hooper:2017gtd} pointed out that the enhanced flux and small extension of the gamma-ray source around Geminga are likely to be interpreted by a slow-diffusion environment. This view was supported by the subsequent morphology measurement of HAWC \citep{Abeysekara:2017old}. As the gamma-ray profile of the Geminga halo can be fitted well by the diffusion model with a small diffusion coefficient, the slow-diffusion picture has been widely accepted. However, it is challenging to interpret the origin of the slow-diffusion environment.

\subsection{Self-generated MHD Turbulence}
\label{subsec:self}
\begin{figure}[t!]
\begin{center}
\includegraphics[width=8cm]{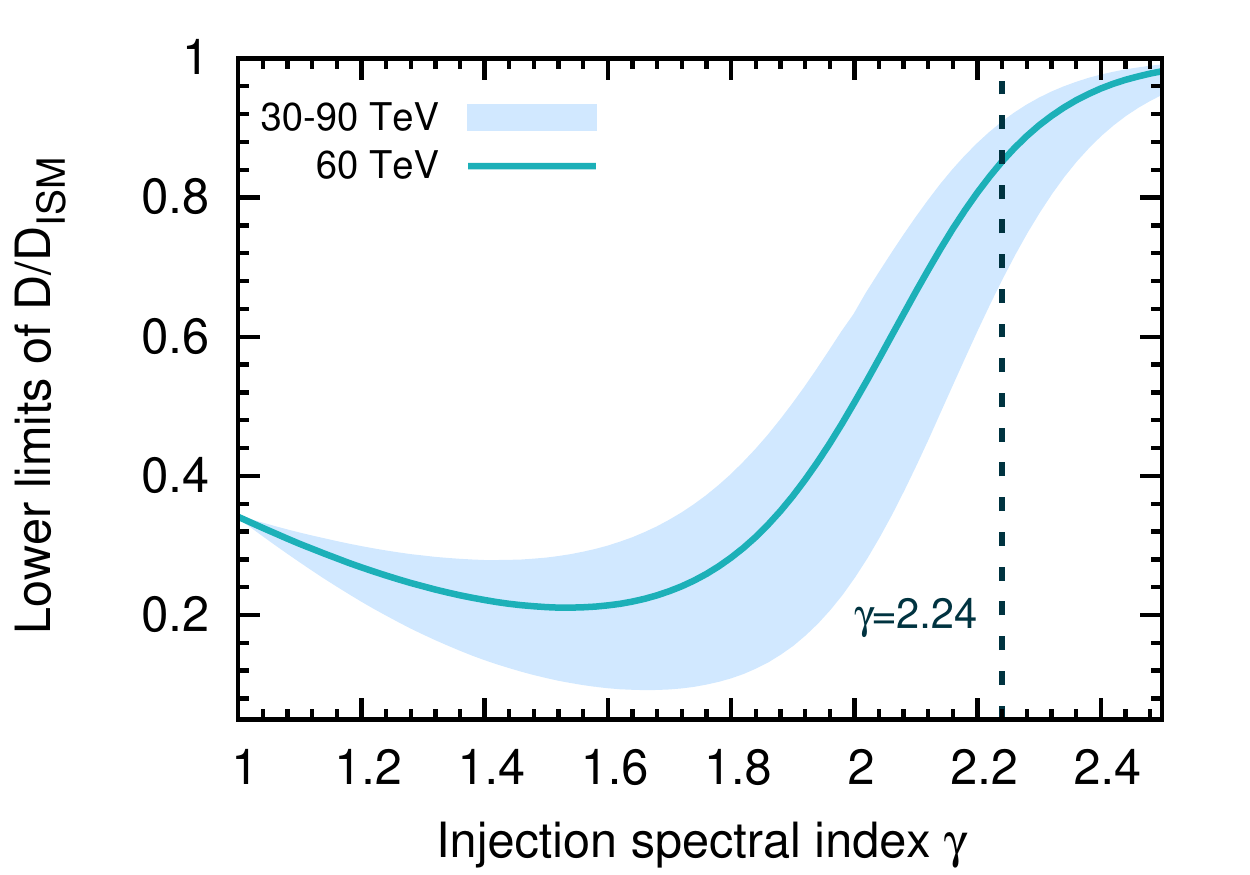}
\includegraphics[width=8.3cm]{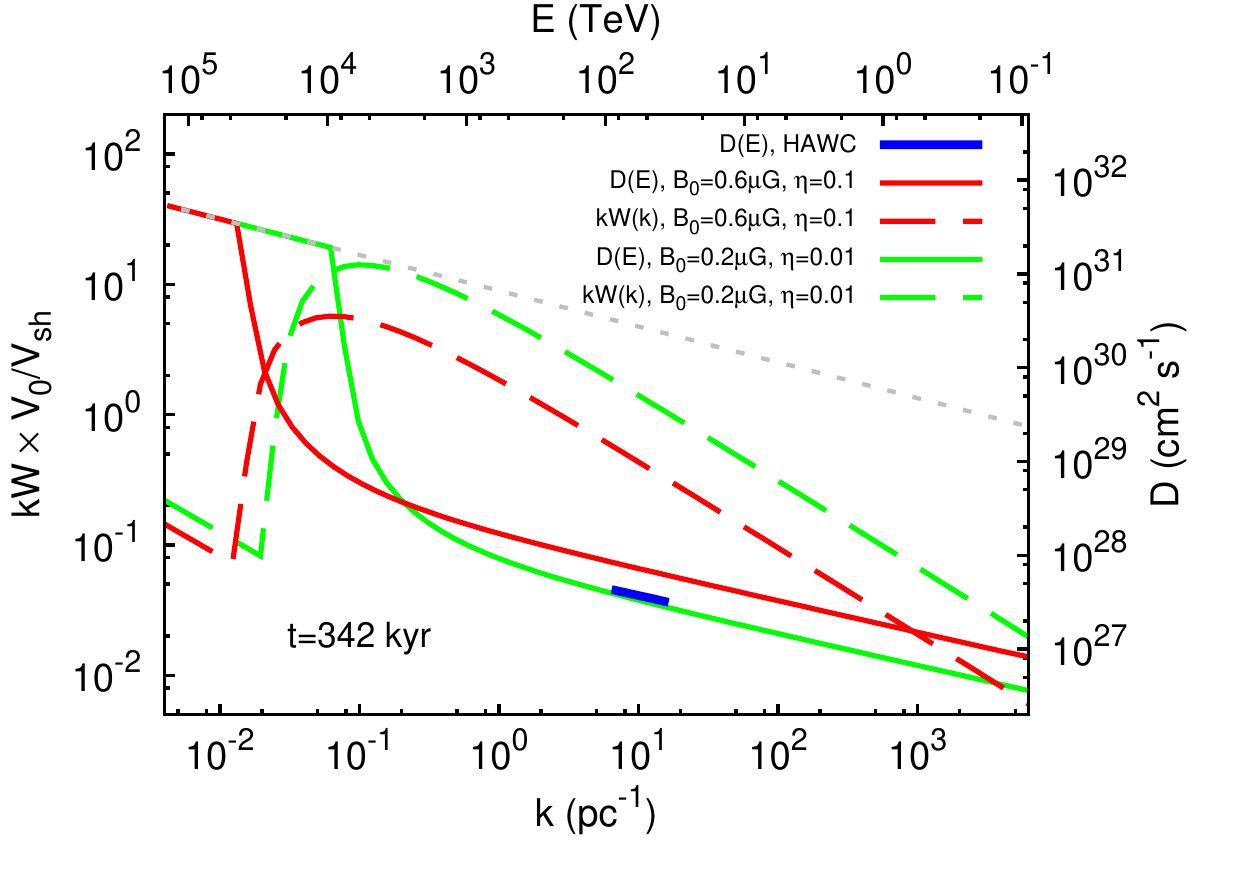}
\end{center}
\caption{\textbf{Left:} Lower limit of the diffusion coefficient in the Geminga halo as a function of the injection spectral index under the self-generated scenario. The RSI is assumed to account for the turbulent growth. One-dimensional diffusion is assumed. Energy losses of electrons and turbulent damping are ignored, which are optimistic conditions for turbulent growth. Electrons injected during the last third of the age of Geminga are all assumed to contribute to the turbulent growth. \textbf{Right:} Wave-number spectrum of the MHD turbulence and the corresponding diffusion coefficient for the Geminga halo, assuming the turbulence is provided by the SNR shock. $B_0$ is the background magnetic field strength, and $\eta$ is the conversion efficiency from the SNR initial energy to the MHD turbulenct energy. The gray dotted line is the average diffusion coefficient in the Galaxy. Both figures are taken from \citet{Kun:2019sks}.
}
\label{fig:origin}
\end{figure}

The most straightforward interpretation may be the self-generated scenario, which has been adopted to predict the slow-diffusion environment around SNRs \citep{Ptuskin:2008zz,Malkov:2012qd,DAngelo:2015cfw,Recchia:2021vfw} and spatially dependent diffusion of CRs at the Galactic scale \citep{Recchia:2016bnd,Evoli:2018nmb}. A large density gradient of CRs near the sources can induce the growth of the small-scale magnetohydrodynamic (MHD) turbulence of the background plasma, known as the resonant streaming instability (RSI) \citep{1971ApJ...170..265S}. CRs can then be trapped by the enhanced MHD turbulence generated by themselves, corresponding to the suppression of the diffusion coefficient. In the one-dimensional (1D) diffusion case, numerical calculations show that the positron/electron pairs released by the Geminga PWN may suppress the diffusion coefficient to the observed level through RSI if the pulsar motion is ignored \citep{Evoli:2018aza,Kun:2019sks,Mukhopadhyay:2021dyh}. However, the proper motion measurement of Geminga indicates that the pulsar has been traveling about $\ang{17}$ from its birthplace \citep{2007Ap&SS.308..225F}, significantly larger than the size of the pulsar halo. It means that the positron/electron pairs released at the early age of Geminga cannot contribute to the formation of the slow-diffusion zone. \citet{Kun:2019sks} proved through a simple analytical derivation that if the pulsar motion is taken into account, the diffusion coefficient cannot be suppressed to the required level even under very optimistic conditions for the RSI growth (see the left of Figure~\ref{fig:origin}). Moreover, the 1D diffusion mode may eventually evolve into the 3D mode as suggested by the symmetry of the Geminga halo. \citet{Schroer:2020dqy} also pointed out that the 1D flux-tube assumption will be broken by the transverse expansion caused by the transverse pressure gradient. The RSI growth is further limited in the 3D mode \citep{Mukhopadhyay:2021dyh} as the injected electrons are diluted compared with the 1D case.

Another branch of the self-generated scenario is the turbulent growth through the nonresonant streaming instability (NRSI) \citep{2004MNRAS.353..550B}. The growth rate of NRSI could be larger than the NRI \citep{Schroer:2020dqy}, while it is proportional to the total CR current, which is zero for the case of the positron/electron pairs. Although the possible asymmetry between positrons and electrons during the acceleration or escape process may excite the NRSI \citep{Schroer:2022gau}, quantitative calculations are needed to test this interpretation.

\subsection{External Sources of MHD Turbulence}
\label{subsec:external}
Besides being excited by pulsars themselves, the slow-diffusion environment could also be generated by external sources. Stellar feedback, such as SN explosion or stellar wind, is believed to be the main source of turbulence in the ISM \citep{Ferriere:2019iqq}. For pulsars, their parent SNRs could be the most straightforward sources for the turbulent environment. In the SNR shock frame, the upstream plasma loses part of its kinetic energy when streaming through the shock front, and this part of the energy is transferred into turbulence and thermal energy behind the shock \citep{Bell:1978zc}. Thus, the downstream region could be highly turbulent \citep{2007ApJ...663L..41G}, although the turbulence will gradually decrease with the SNR evolution. Assuming the MHD turbulence is injected at 10~pc and cascades to smaller scales, \citet{Kun:2019sks} pointed out that the diffusion coefficient downstream of the SNR shock could be suppressed by more than two orders of magnitude if 1-10\% of the initial energy of SNR is converted to the MHD turbulence (see the right of Figure~\ref{fig:origin}). If the pulsars are still inside their associated SNRs, the observed slow-diffusion environment could be understood.

For Geminga, the parent SNR has not been observed, possibly owing to old age. The proper motion of Geminga indicates that it has been $\approx70$~pc away from the SNR center, so there are works considering that Geminga has already left behind the SNR. However, owing to the rarefied circumstance of Geminga \citep{2003Sci...301.1345C}, the scale of the SNR may reach $\sim100$~pc \citep{Leahy:2017nrs,Kun:2019sks}, which means that Geminga may still be downstream of the SNR shock. This scenario is consistent with the high-ionization degree environment of Geminga indicated by the measurement of the H$\alpha$ luminosity \citep{2003Sci...301.1345C}. Furthermore, as the speed of sound in the outer part of the SNR is significantly smaller than that of the motion of Geminga, this scenario is consistent with the bow-shock nature of the Geminga PWN, meeting CT~3 given in Section~\ref{subsec:CTs}. For the other known pulsar halos, Monogem is likely inside its parent SNR, namely the Monogem ring \citep{1996ApJ...463..224P,2018MNRAS.477.4414K}, which is still observable in X-ray owing to its younger age. However, the SNRs of PSR~J0622$+$3749 and PSR~J1831$+$0952 have not been detected. Searching for traces of the associated SNRs may be an important subject for this interpretation.

On the other hand, a slow-diffusion environment could also exist upstream of the SNR shock, as mentioned in Section~\ref{subsec:self}. This scenario is supported by the observations of the interaction between the escaping CR nuclei from SNRs and the molecular clouds near the SNRs \citep{Fujita:2009ak,Li:2011qx}. Thus, pulsar halos may be generated if the host pulsars are embedded in the turbulent medium upstream of the SNR shocks. Unlike positron/electron pairs discussed in Section \ref{subsec:self}, CR nuclei could effectively induce the NRSI, and the slow-diffusion zone may be retained for a long time owing to the large growth rate of NRSI \citep{Schroer:2020dqy,Schroer:2022gau}.

\citet{Kun:2019sks} also proposed another possible scenario: Geminga is inside a stellar-wind bubble, the Gemini H$\alpha$ Ring \citep{2007ApJ...665L.139K,2018MNRAS.477.4414K}. The shocked wind may provide a fresh MHD turbulence environment for Geminga. The Gemini H$\alpha$ Ring is likely generated by several OB-type stars that may have similar distances with Geminga \citep{2018MNRAS.477.4414K}. Further quantitative calculations are needed to test this scenario.

\subsection{Anisotropic Diffusion}
\label{subsec:anisotropy}
Some intriguing models could interpret the pulsar halos without assuming anomalously slow diffusion. The typical coherent length of the magnetic field in the ISM is $\sim50-100$~pc, and CRs should mainly propagate along the mean magnetic field lines in the case of Alfv\'enic turbulence \citep{Goldreich:1994zz}. Meanwhile, CRs diffuse perpendicularly to the mean magnetic field with a much smaller diffusion coefficient of $D_\perp=D_\parallel M_A^4$, where $D_\parallel$ is the parallel diffusion coefficient assumed to be the average value in the Galaxy, and $M_A$ is the Alfv\'{e}nic Mach number \citep{Yan:2007uc}. The CR propagation, in this case, is anisotropic, which seems to be inconsistent with the symmetry of the pulsar halos. However, \citet{Liu:2019zyj} pointed out that if the local interstellar magnetic field of the Geminga pulsar is aligned with our line of sight towards the pulsar, the morphology of the pulsar halo could be interpreted by the inefficient perpendicular diffusion assuming $M_A=0.2$. If so, sources of strong MHD turbulence discussed in Section \ref{subsec:self} and \ref{subsec:external} are not required. Another advantage of this model is that it could naturally interpret the weak X-ray synchrotron emission in the vicinity of Geminga \citep{Liu:2019sfl}.

Obviously, this model has stringent requirements for the inclination angle $\phi_{\rm incl}$ between the line of sight and the direction of the magnetic field tube the pulsar embedding. A larger $M_A$ can reduce asymmetry but will boost the extension of the source at the same time. The calculation of \citet{2022arXiv220508544D} indicates that to consistently reproduce the symmetry and extension of the Geminga halo, the permitted region in the $M_A-\phi_{\rm incl}$ plane is very small, which means that the possibility of finding multiple Geminga-type halos is extremely low. This scenario may predict more pulsar halos with significant asymmetry, which could be tested by future experiments \citep{2022arXiv220514563Y}.

\begin{figure}[t!]
\begin{center}
\includegraphics[width=8cm]{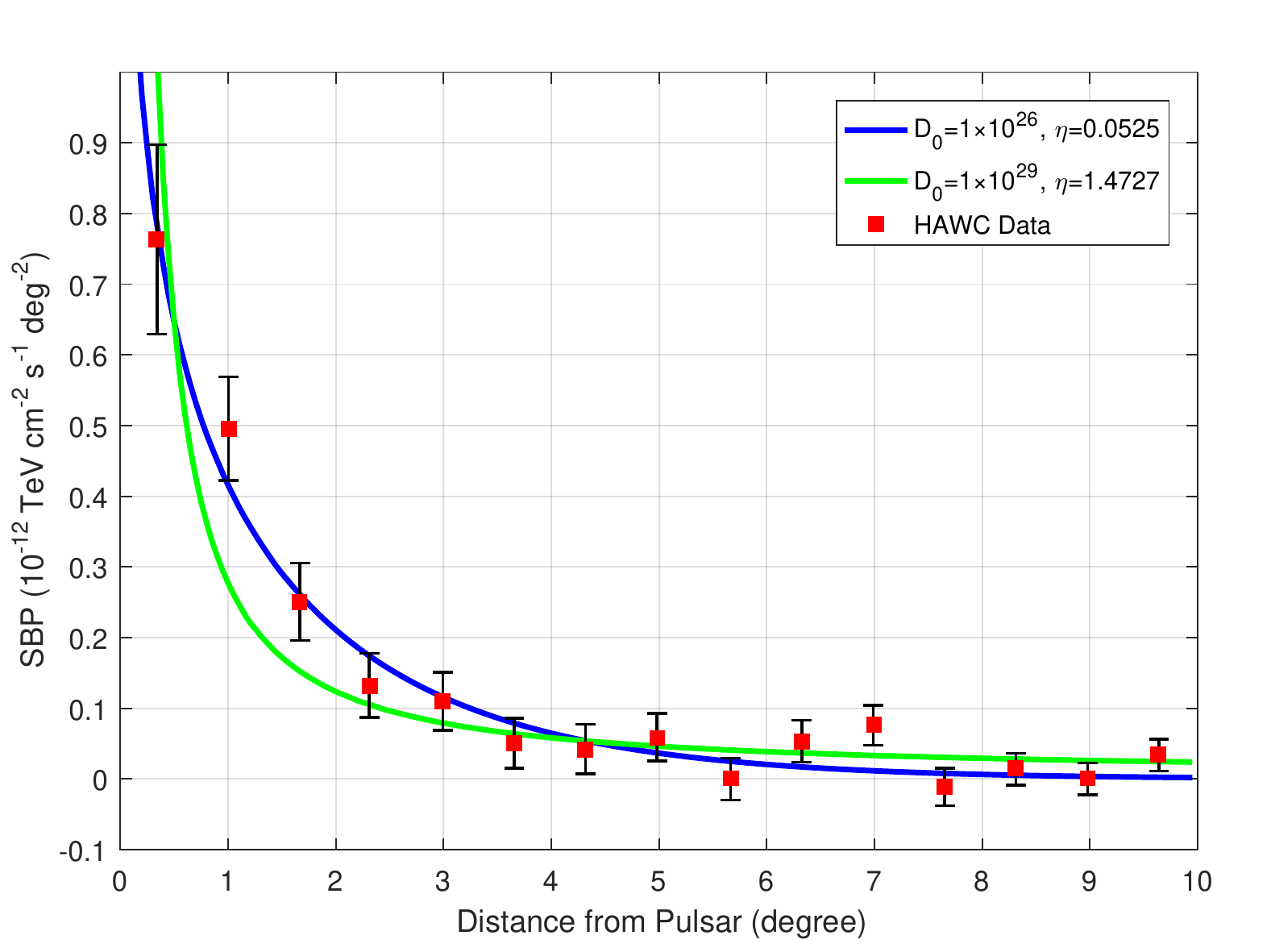}
\includegraphics[width=8cm]{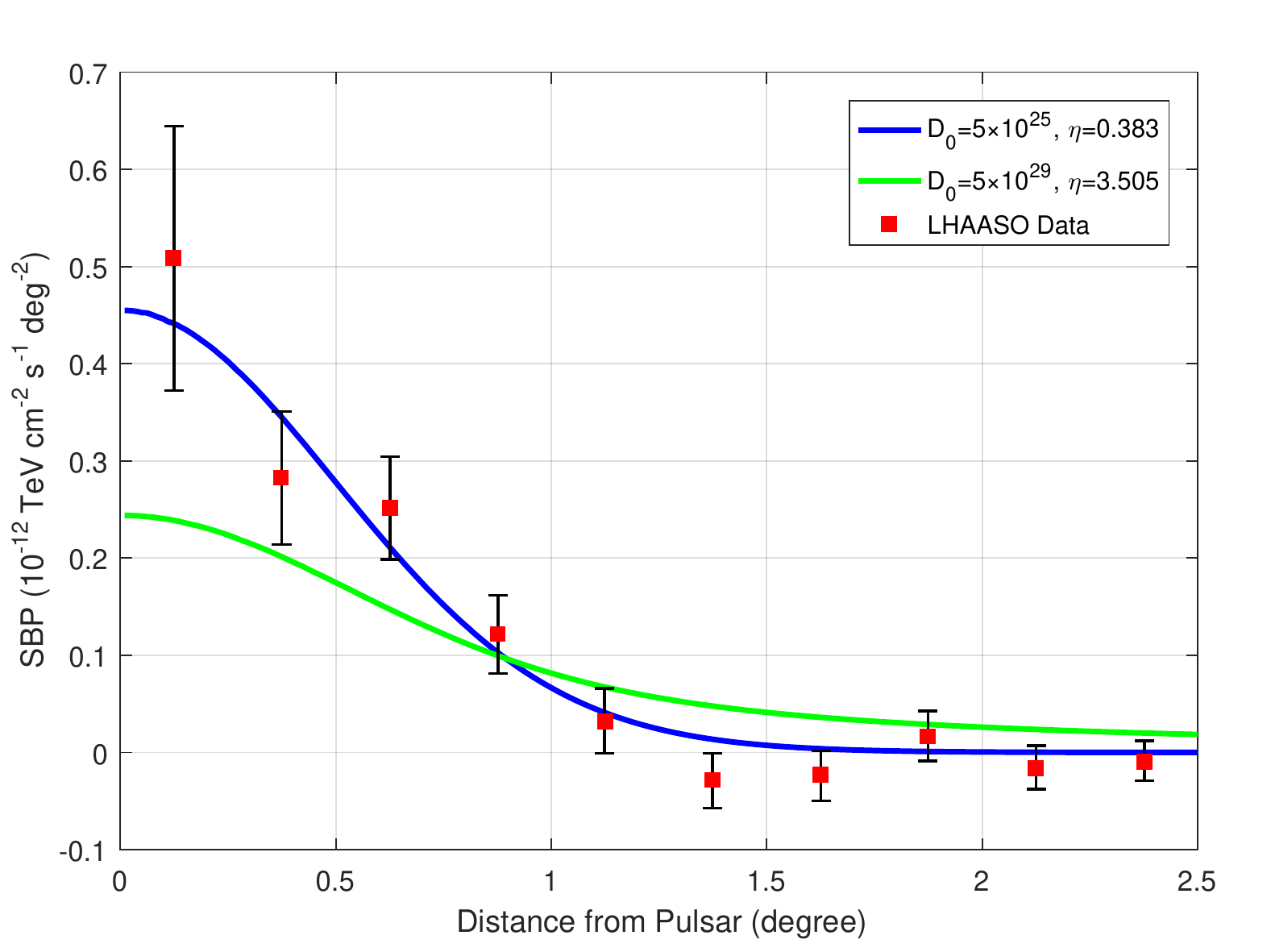}
\end{center}
\caption{Gamma-ray profiles of the Geminga halo (left) and LHAASO J0621$+$3755 (right) after the relativistic correction to the propagation equation, for both the fast- and slow-diffusion scenarios. $D_0$ is the diffusion coeffcient at 1~GeV, and $\eta$ is the conversion efficiency from the pulsar spin-down energy to the electron injection energy. Both figures are taken from \citet{Bao:2021hey}.}
\label{fig:profiles}
\end{figure}

\subsection{Relativistic Diffusion}
\label{subsec:relativistic}
The commonly used diffusion model for CR propagation has the superluminal problem. This problem can be ignored if the typical diffusion scale ($\sim\sqrt{Dt}$, where $D$ is the diffusion coefficient) is significantly smaller than the distance traveled at the speed of light ($ct$), namely $t\gg D/c^2$. Otherwise, the diffusion model will predict a significantly broader CR distribution than the real case, and a relativistic correction to the solution is essential \citep{Dunkel:2006kn,Aloisio:2008tx}. Another way to describe this issue is that if $ct$ is smaller than the mean free path of CRs, $\lambda$, that defines the diffusion coefficient with $D=1/3\lambda c$, the CR propagation should be ballistic rather than diffusive. The critical time is also $t\sim D/c^2$. It can be seen that the effect is more significant for a larger $D$ as the distribution of CRs injected in a longer period needs to be corrected in this case. Taking this effect into account, the calculation of \citet{Recchia:2021kty} showed that a steep gamma-ray profile around Geminga is also predicted assuming the average diffusion coefficient in the Galaxy, and a slow-diffusion environment may no longer be required. This scenario seems attractive as it could be more general than the geometric effect introduced in Section \ref{subsec:anisotropy}.

However, two problems remain to be solved for the fast diffusion model: the goodness of fit and the conversion efficiency. For the Geminga halo, the minimal reduced $\chi^2$ of the profile fit is $\approx2$, and the required conversion efficiency is $\approx150\%$. As mentioned in Section \ref{subsec:CTs}, a reasonable model of pulsar halo should interpret the observations with a conversion efficiency smaller than 100\%. These problems are more severe for LHAASO~J0621$+$3755 as shown by the calculation of \citet{Bao:2021hey}: the minimal reduced $\chi^2$ of the profile fitting is $\approx4$, corresponding to exclusion with a confidence level of $99.996\%$, and the required conversion efficiency is $\approx350\%$\footnote{For LHAASO~J0621$+$3755, \citet{Recchia:2021kty} obtained different results from \citet{Bao:2021hey} as they took a 1D Gaussian function as the point-spread-function (PSF). However, the PSF should be defined in the 2D plane.}. By comparison, the slow-diffusion model is hardly affected by the relativistic correction. The reduced $\chi^2$ for both the sources are around $1$, and the conversion efficiencies are significantly smaller than $100\%$. Thus, the slow-diffusion model is obviously the more reasonable one \citep{Bao:2021hey}. The profile fits of the two pulsar halos are shown in Figure~\ref{fig:profiles} for both the fast- and slow-diffusion models. 

\section{Pulsar Halos as Indicators of Galactic CR Propagation}
\label{sec:prop}
Pulsar halos could be the most potent indicator of CR propagation in the localized medium. Molecular clouds illuminated by escaping CRs from SNRs can also measure the diffusion coefficient in the near-source region \citep{Fujita:2009ak,Li:2011qx}. However, this method is difficult to measure the CR spatial distributions in the ISM and investigate detailed mechanisms of CR propagation. In comparison, pulsar halos' morphology can straightforwardly indicate the propagation of CR electrons in the ISM. We will show below that the gamma-ray spectra of pulsar halos are also valuable for constraining propagation models. The measurements with pulsar halos are complementary to the global probes of Galactic CR propagation.

\subsection{Local Propagation}
\label{subsec:prop_local}
Although the diffusion model is quite simple, it successfully describes CR propagation in the Galaxy \citep{1964ocr..book.....G}. The Brownian motion could simulate particle transport in the turbulent interstellar magnetic field on microscopical scales. However, the normal diffusion model is applicable only when the inhomogeneity of the ISM is negligible for the scale of interest. The inhomogeneity of the ISM can have two types of effects on diffusion properties \citep{1999cs..book.....U}:

\begin{enumerate}
 \item The diffusion coefficient is spatially dependent, which can be
considered as a superposition of normal diffusion processes.
 \item The particle motion is different from the Brownian motion, and the shape of the diffusion packet is no longer Gaussian, known as the anomalous diffusion.
\end{enumerate}

The extremely small diffusion coefficient measured with the pulsar halos cannot be representative in the Galaxy. To reconcile the slow diffusion around pulsars and the typical diffusion coefficient of the Galaxy, the two-zone diffusion model is proposed \citep{Hooper:2017gtd,Fang:2018qco}, where the diffusion coefficient is only suppressed within $r_\star$ around pulsars and recovers to the typical value outside $r_\star$. For the mechanisms discussed in Sections \ref{subsec:self} and \ref{subsec:external}, the slow diffusion could indeed be a near-source phenomenon. The self-excited mechanism may only suppress the diffusion coefficient within $\sim50$~pc around the pulsar \citep{Evoli:2018aza}, as the particle gradient at farther distances is too low to induce turbulent growth. If the slow-diffusion environment is produced by the parent SNR, $r_\star$ could be comparable to the SNR size and may also be $\sim50$~pc. The two-zone diffusion belongs to the first scenario listed above.

\begin{figure}[t!]
\begin{center}
\includegraphics[width=8.0cm]{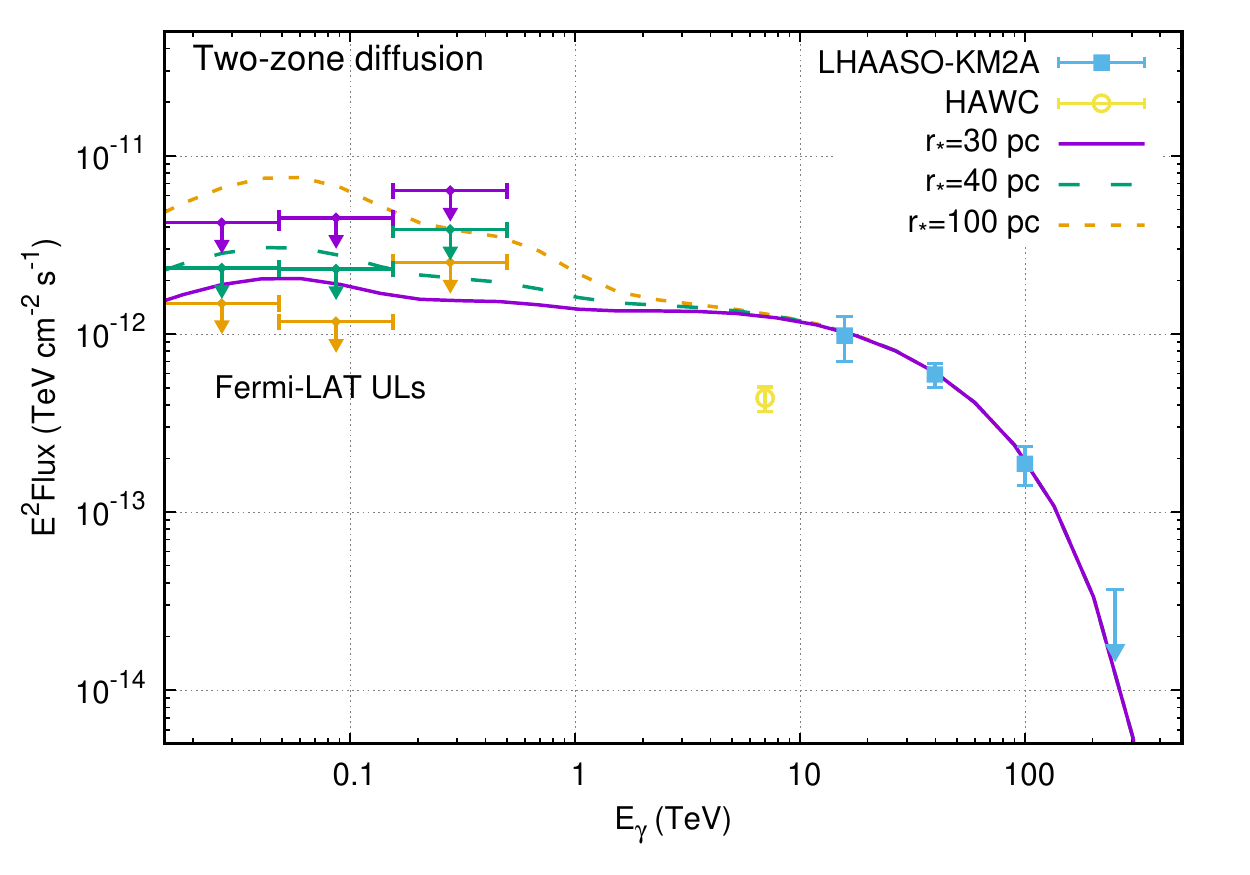}
\includegraphics[width=8.0cm]{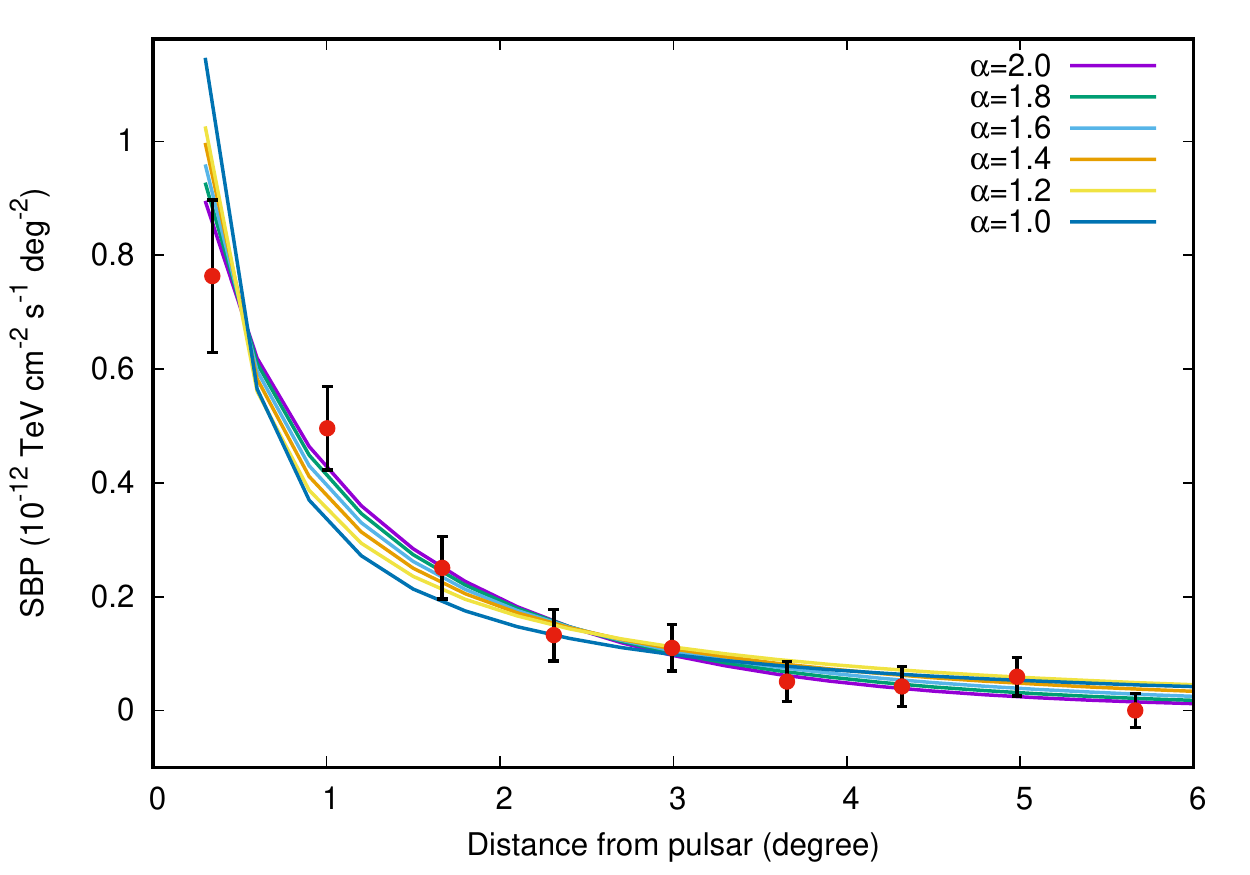}
\end{center}
\caption{\textbf{Left:} Interpretation of the wide-band gamma-ray spectrum of LHAASO~J0621$+$3755 with two-zone diffusion models \citep{Fang:2021qon}. Different sizes of the slow-diffusion zone are adopted. The Fermi-LAT ULs vary with the models. \textbf{Right:} Fitting results of the Geminga halo's gamma-ray profile measured by HAWC with superdiffusion models \citep{Wang:2021xph}, where $\alpha$ is the superdiffusion index. Superdiffusion degenerates to the normal diffusion when $\alpha=2$.}
\label{fig:prop}
\end{figure}

Gamma-ray spectra of pulsar halos suggest the possibility of two-zone diffusion. For LHAASO~J0621$+$3755, the TeV spectrum measured by the 1.3~km$^2$ array of LHAASO (LHAASO-KM2A) and the GeV upper limits (ULs) given by the Fermi Large Area Telescope (Fermi-LAT) cannot be interpreted consistently with the one-zone normal diffusion model, or an unreasonably hard injection spectrum is needed \citep{Aharonian:2021jtz}. We continue the topic of the injection spectrum in Section~\ref{sec:injection}. Under the two-zone diffusion assumption, electrons with lower energies have a longer lifetime and have more chance to escape from the slow-diffusion zone, which can suppress the low-energy spectrum below the Fermi-LAT ULs, as shown in the left of Figure~\ref{fig:prop}. Thus, the two-zone diffusion model can interpret the wide-band gamma-ray spectrum of LHAASO~J0621$+$3755 with a reasonable injection spectrum if $r_\star\lesssim50$~pc \citep{Fang:2021qon}, consistent with the theoretical expectations. For the Geminga halo and HESS~J1831$-$098, the two-zone diffusion model may also help to interpret the low flux or non-detection in the GeV band \citep{Fang:2022mdg,2022arXiv220713533F}.

On the other hand, the inhomogeneity of the ISM may arise on all scales. In this case, the CR propagation cannot be described only by the spatially dependent diffusion. The ISM is more likely to be a fractal type, and the normal diffusion can be generalized to superdiffusion \citep{2001NuPhS..97..267L,2003NIMPB.201..212L,2012GrCo...18..122U,2017arXiv170306486U}. Microscopically, the particle motion is described by L\'evy flight instead of the Brownian motion, where occasional very long steps are permitted. The widening of the diffusion packet with time is faster for L\'evy flight than the normal diffusion case; hence superdiffusion is named. The fractional Laplacian operator, $\Delta^{\alpha/2}$, is used for the superdiffusion equation. The index $\alpha$ represents the degree of fractality of the ISM, which is defined in $(0,2]$. When $\alpha=2$, the propagation degenerates to the normal diffusion. Superdiffusion has been observed in interplanetary space \citep{2009ApJ...693L.118P,2009AdSpR..44..465P}, the ISM \citep{2016A&A...596A..34P}, and the cluster of galaxies \citep{1997A&A...327..432R}, indicating that it could be common in astrophysics. This picture is also adopted in the global propagation of Galactic CRs to interpret features of the CR energy spectra \citep{2001NuPhS..97..267L,2013JPhCS.409a2042V}.

As superdiffusion yields different CR spatial distributions from the normal diffusion, the morphology of pulsar halos can be used to test the superdiffusion model. The Geminga halo is the best target among the known pulsar halos. Owing to the large extension of the Geminga halo, the morphology measurements are affected little by the resolution of the experiments. \citet{Wang:2021xph} fitted the profile of the Geminga halo measured by HAWC with different $\alpha$ and found that a model with $\alpha\lesssim1.3$ is disfavored at 95\% confidence level. The fitting results are shown in the right of Figure~\ref{fig:prop}. Compared with the normal diffusion model, profiles predicted by superdiffusion models are steeper near the source while flatter away from the source. Models with $\alpha$ significantly smaller than $2$ give poor fits to the near-source data, while the measurement still allows superdiffusion with $\alpha\lesssim2$. \citet{Fang:2021qon} pointed out that superdiffusion is also helpful in interpreting the non-detection of the GeV emission of LHAASO~J0621$+$3755.

\subsection{Global Propagation}
\label{subsec:prop_global}
The CR boron-to-carbon ratio (B/C) is the major probe of the global propagation of Galactic CRs \citep{Strong:2007nh}. The CR B is entirely secondary, produced by the interactions between the CR C, N, and O and the ISM. Thus, B/C is directly proportional to the thickness of the Galactic diffusion halo and inversely proportional to the diffusion coefficient. Ratios between unstable secondary to primary CR nuclei, such as $^{10}$Be/$^{9}$Be, can further help to disentangle the diffusion coefficient from the halo thickness. The average diffusion coefficient in the Galaxy is then obtained.

However, the diffusion coefficient is very likely spatially dependent in the Galaxy as the distribution of turbulent sources is not homogeneous. This picture is supported by observations. The measured turbulent magnetic field energy in the inner Galactic disk could be one order of magnitude larger than that in the high-latitude region of the outer Galaxy \citep{2017ARA&A..55..111H}. Spatially dependent propagation models can also interpret various features of CR measurements, such as the spectral hardening of the nuclei spectra, the radial distributions of the CR proton densities and spectral indices, the diffuse gamma-ray spectrum in different sky regions, and the CR anisotropy \citep{Tomassetti:2012ga,2016PhRvD..94l3007F,2018PhRvD..97f3008G,Zhao:2021yzf}. Among these measurements, the CR anisotropy is sensitive to the diffusion coefficient in the nearby ISM. Based on the assumption that the 10~TeV spectral bump of the CR proton spectrum is due to a nearby CR source \citep{DAMPE:2019gys,Yue:2019sxt,Yuan:2020ciu}, \citet{Fang:2020cru} argued that a slow-diffusion region could exist between this nearby source and the solar system to consistently interpret the proton spectral feature and the low CR anisotropy. Besides, spatially dependent diffusion is predicted by different theoretical models \citep{Evoli:2013lma,Evoli:2018nmb}.

The discovery of pulsar halos is direct evidence of the spatially dependent diffusion in the Galaxy if the morphology of pulsar halos is indeed due to slow diffusion. 
For the mechanisms introduced in Sections~\ref{subsec:self} and \ref{subsec:external} (except for the stellar-wind bubble case), we may use the explosion rate of supernovae to estimate the birth rate of the slow-diffusion zone. For an average size of $50$~pc and a persisting time of $10^6$~yr, the slow-diffusion regions may occupy $\sim3\%$ of the volume of the Galactic disk \citep{Hooper:2017gtd}. The diffusion coefficient is inversely proportional to the wave-number spectrum of the  MHD turbulence \citep{1971ApJ...170..265S}. Thus, if the wave-number spectrum is 300 times stronger in the slow-diffusion zones, the weight-averaged spectrum in the Galactic disk will be amplified by $\sim10$ times ($300\times3\%$). It corresponds to 10 times suppression of the diffusion coefficient in the Galactic disk on average, which may be consistent with the results implied by the observations mentioned above. However, the universality of pulsar halos is still in doubt, and we will discuss this topic in Section~\ref{subsubsec:universal}.

\section{Pulsar Halos as Indicators of Electron Injection from Pulsar Wind Nebulae}
\label{sec:injection}
There are abundant measurements for the electromagnetic radiation spectrum of PWNe \citep{2017SSRv..207..175R}, and we can infer the electron energy spectrum in PWNe from these measurements. However, as the electron injection spectrum from PWNe may be different from the spectrum in PWNe, we cannot only rely on the observations of PWNe themselves to study the electron escape from PWNe. As pulsar halos are generated by escaping electrons, their spectra are ideal indicators of electron injection. 

\begin{figure}[t!]
\begin{center}
\includegraphics[width=8.0cm]{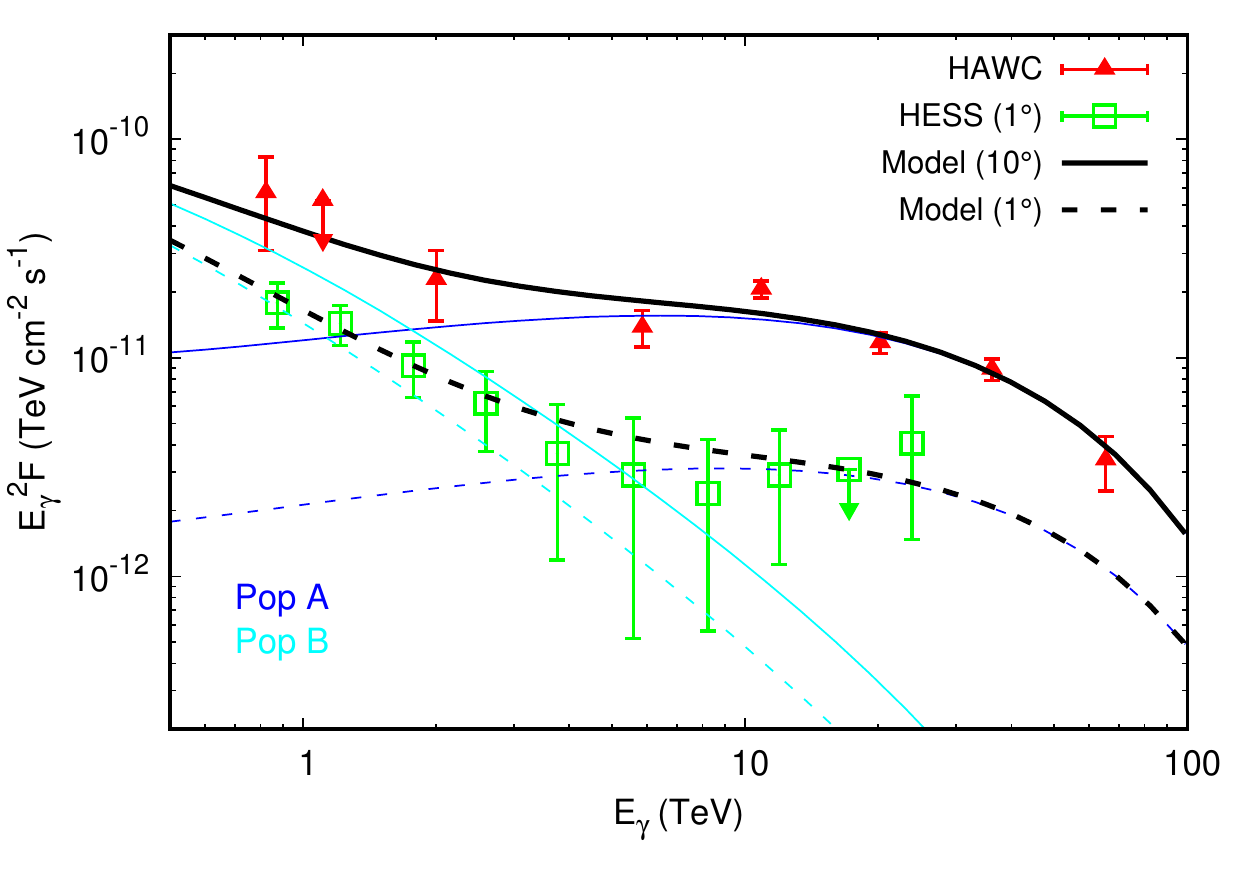}
\includegraphics[width=8.0cm]{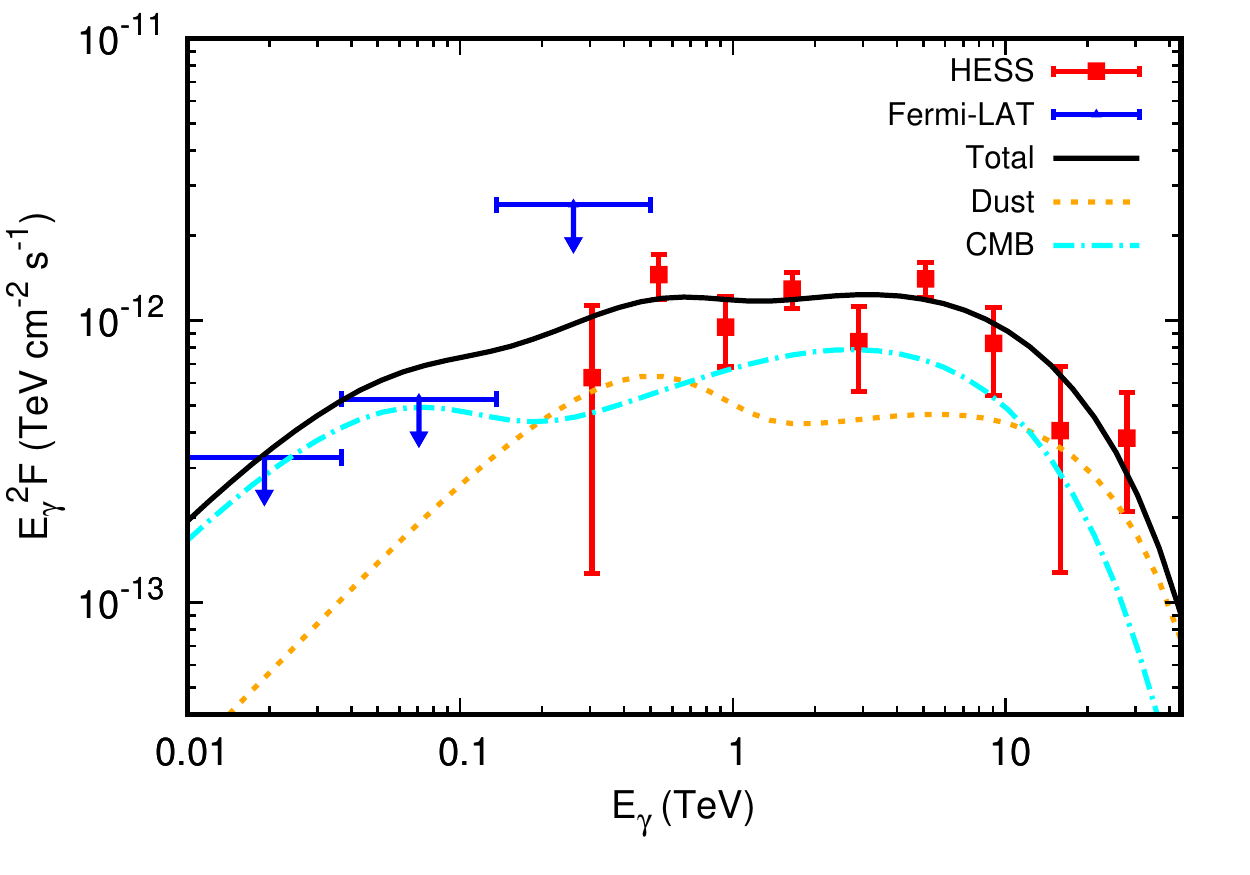}
\end{center}
\caption{\textbf{Left:} Best-fit gamma-ray spectra to the HAWC and H.E.S.S. observations of the Geminga halo with the two-population injection model \citep{Fang:2022mdg}. Pop A represents the freshly accelerated electrons that escape from the PWN through rapid outflows, while Pop B represents the electrons trapped longer in the PWN before escaping. The black solid (dotted) line is the total spectrum within \ang{10} (\ang{1}) around the pulsar, which fits the HAWC (H.E.S.S.) spectrum. \textbf{Right:} Best-fit gamma-ray spectra to the H.E.S.S. data of HESS~J1831$-$098 with a single ECPL injection model \citep{2022arXiv220713533F}. The two sub-components of scattering the dust photons and CMB photons are also separately shown. The Fermi-LAT ULs are shown for comparison but not used in the fitting procedure.}
\label{fig:injection}
\end{figure}

The latest observations from the HAWC and H.E.S.S. experiments exhibit complex features in the gamma-ray spectrum of the Geminga halo \citep{HAWC:2021wia,Mitchell:2021tig}, which is likely attributed to the electron injection process of the Geminga PWN \citep{Fang:2022mdg}. The HAWC spectrum indicates a possible bump feature around 10~TeV. Meanwhile, the
spectrum unexpectedly climbs again below $\approx3$~TeV. This low-energy feature can also be clearly seen from the H.E.S.S. spectrum. These results cannot be interpreted by a single power-law injection spectrum \citep{Fang:2022mdg}. Noting that the X-ray PWN and the TeV gamma-ray halo are generated by electrons with almost the same energy range, \citep{Fang:2022mdg} propose a two-population injection model based on the image and spectral measurements of the Geminga X-ray PWN. One population is the freshly accelerated electrons that escape from the PWN through rapid outflows, corresponding to the lateral tails of the X-ray PWN. This population has an exponentially-cutoff power-law (ECPL) injection spectrum and accounts for the high-energy bump structure of the gamma-ray spectrum. The hard power-law term is suggested by the X-ray PWN spectrum. The high-energy cutoff term corresponds to the acceleration limit of the PWN, and the cutoff energy is $\approx140$~TeV as determined by the data fit. The other population is the electrons trapped longer in the PWN before escaping, corresponding to the axial tail of the X-ray PWN. This population has an additional spectral break in the injection spectrum owing to the synchrotron energy loss of electrons inside the PWN, dominating the low-energy gamma-ray spectrum. The two-population model can interpret the HAWC and H.E.S.S. data well, as shown in the left of Figure~\ref{fig:injection}.

The current gamma-ray spectrum measurements of the Monogem halo and LHAASO~J0621$+$3755 are mainly above $10$~TeV \citep{Abeysekara:2017old,Aharonian:2021jtz}, where the spectra are probably dominated by the high-energy cutoff term mentioned above. While for HESS~J1831$-$098, the gamma-ray spectrum has been measured by H.E.S.S. in the energy range of $\sim0.3-30$~TeV \citep{Sheidaei:2011vg}, which may provide us a comprehensive understanding of its electron injection spectrum. \citet{2022arXiv220713533F} perform a combined fit to the gamma-ray spectrum and morphology of HESS~J1831$-$098, finding the electron injection spectrum can be described by the ECPL form well. The best-fit spectrum is shown in the right of Figure~\ref{fig:injection}. The best-fit cutoff energy is $52$~TeV, while the best-fit power-law index is $0.88$, significantly smaller than the typical value of PWNe ($1.5$). The Fermi-LAT flux ULs at lower energies even requires a harder power-law term. Such a hard spectrum could not be the general case as constrained by the CR positron spectrum \citep{2022arXiv220505200B}, although simulations indicate that an electron spectral index of $1.0$ is possible for bow-shock PWNe \citep{Bykov:2017xpo}. The two-zone diffusion model or the time-delayed slow-diffusion model may interpret the H.E.S.S. and Fermi-LAT data with a milder power-law term of the injection spectrum \citep{2022arXiv220713533F}. These models can suppress the low-energy gamma-ray fluxes and could be the more realistic scenarios considering the possible origins of the slow-diffusion zone (Sections~\ref{subsec:self} and \ref{subsec:external}).

\section{Positron excess}
\label{sec:positron}
In early works, CR positrons were considered to be secondary products generated by the collisions between CR nuclei and the interstellar material \citep{Moskalenko:1997gh}. However, with the measurement of the positron energy spectrum breaking through $\sim100$~GeV, secondary positrons alone can no longer interpret the high-energy fluxes, known as the positron excess \citep{PAMELA:2008gwm,Fermi-LAT:2011baq,AMS:2013fma}. This phenomenon 
has attracted a lot of attention as the high-energy positrons may come from the annihilation or decay of dark matter \citep{Bergstrom:2008gr,Cholis:2008hb,Yin:2008bs,Yuan:2013eja}. However, the contribution of dark matter to the positron spectrum is strongly constrained by the gamma-ray observation of Fermi-LAT on the dwarf galaxies and so on \citep{Huang:2012yf,Fermi-LAT:2015att,Lin:2014vja,Liu:2016ngs}. The positron excess could also be originated from nearby pulsars \citep{Hooper:2008kg,Yuksel:2008rf,Yin:2013vaa,DellaTorre:2015xna}. Electron-positron pairs are produced in the magnetosphere of pulsars and further accelerated in the PWNe.

\subsection{Geminga: the Main Candidate}
\label{subsec:geminga}
Geminga is a likely source of the positron excess owing to its proper age and distance and the relatively large spin-down luminosity \citep{Yin:2013vaa}. Moreover, the X-ray PWN and gamma-ray pulsar halo of Geminga indicate that the Geminga PWN can indeed accelerate and release high-energy positrons. 

\begin{figure}[t!]
\begin{center}
\includegraphics[width=8.0cm]{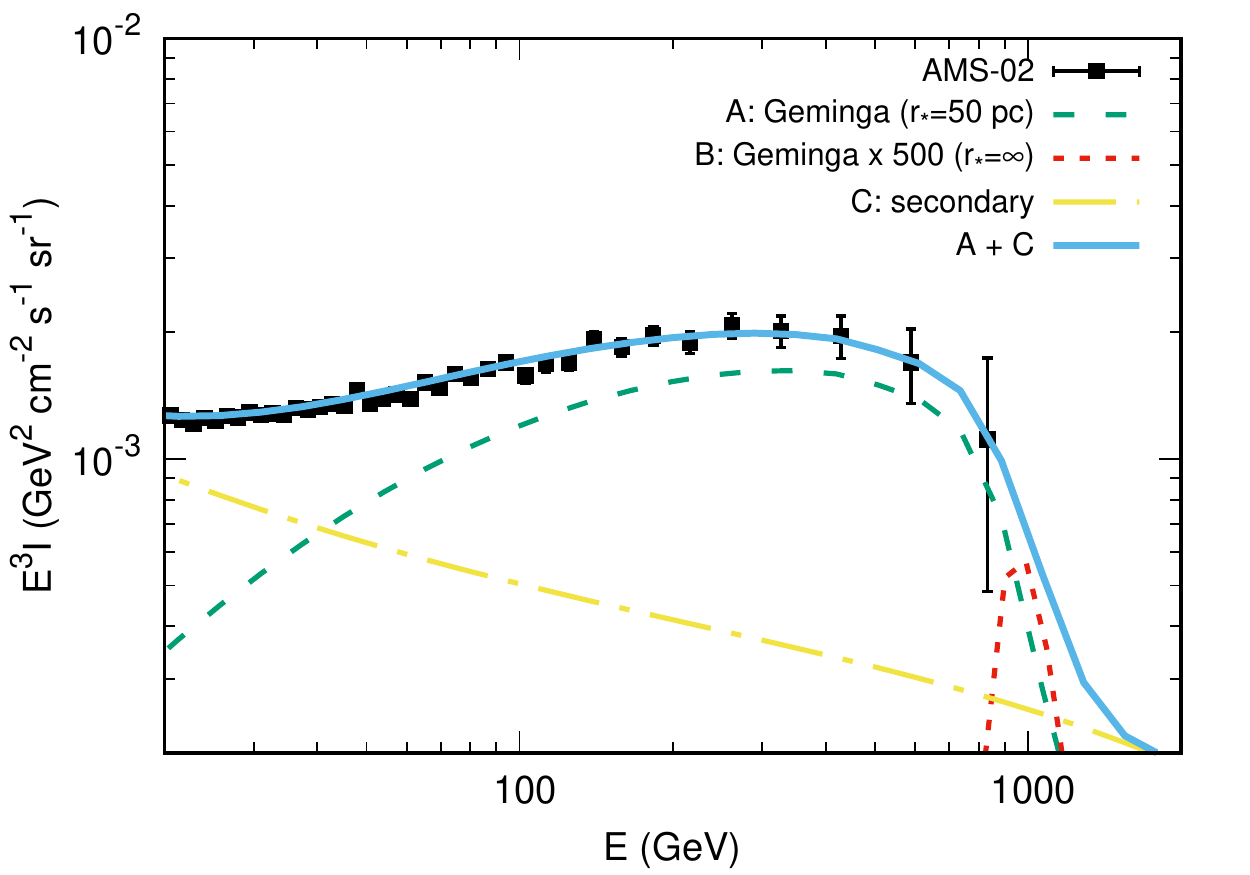}
\includegraphics[height=5.6cm,width=8.0cm]{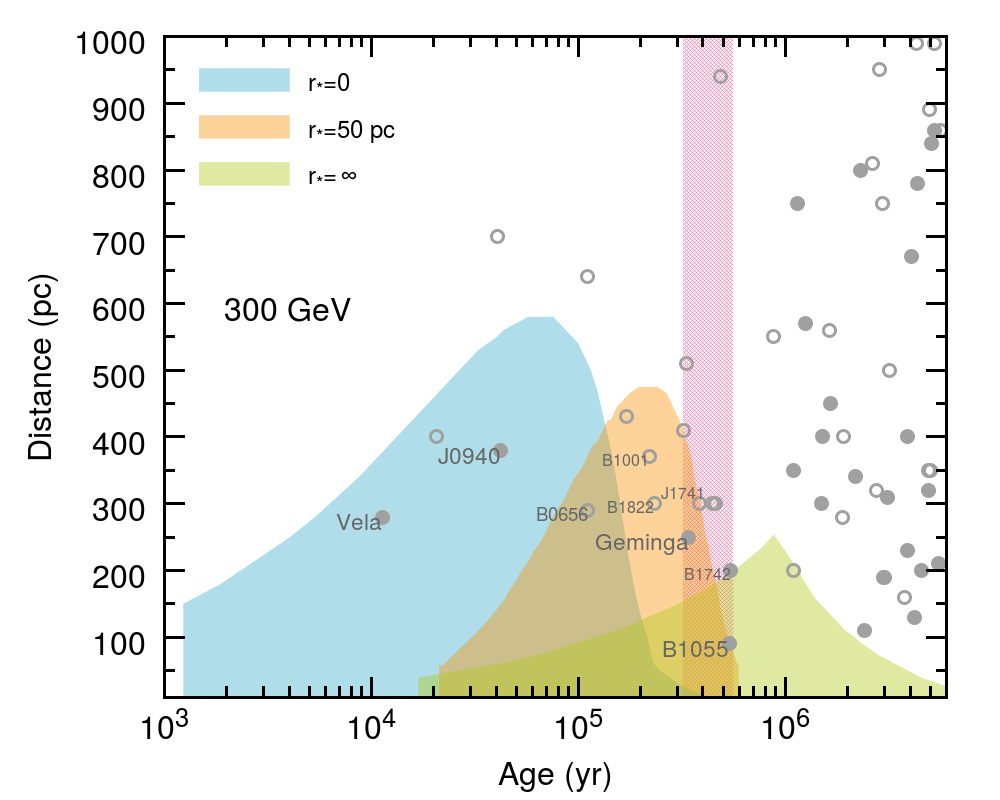}
\end{center}
\caption{\textbf{Left:} Interpretation of the positron excess with Geminga under the two-zone diffusion model ($r_\star=50$~pc). The result of the one-zone slow-diffusion model ($r_\star=\infty$) is also shown for comparison. This figure is adapted from \citet{Fang:2018qco}. \textbf{Right:} Contours of the positron flux at 300~GeV as a function of pulsar age and distance. Pulsars inside the shaded areas are those can contribute more than half of the measured flux of AMS-02 at 300~GeV. Three different models are presented with different colors. The red band indicates the range of the cooling time corresponding to positron spectral cutoff measured by AMS-02 \citep{Aguilar:2019owu}. One may refer to \citet{Fang:2019ayz} for further details of this figure.}
\label{fig:positron}
\end{figure}

If pulsar halos are interpreted by the slow-diffusion scenario discussed in Sections \ref{subsec:self} and \ref{subsec:external}, the positron flux produced by pulsars at Earth will be significantly affected. Assuming the small diffusion coefficient around Geminga can be extrapolated to the whole space between Geminga and the solar system, the positrons produced by Geminga can hardly reach Earth, as shown by the red dotted line in the left of Figure~\ref{fig:positron}. If so, Geminga will be unlikely to account for the positron excess \citep{Abeysekara:2017old}. However, as we have mentioned in Section~\ref{subsec:prop_local}, the slow-zone diffusion zones around pulsars may have a typical size of $\sim50$~pc considering the possible mechanisms, which is also supported by the observations of LHAASO~J0621$+$3755 \citep{Fang:2021qon}. Under the two-zone diffusion model with $r_\star\sim50$~pc, Geminga can still interpret the positron excess well (see the left of Figure~\ref{fig:positron}). The needed conversion efficiency is even more reasonable than the commonly used one-zone fast diffusion model \citep{Fang:2018qco,Profumo:2018fmz,Tang:2018wyr}. Moreover, if superdiffusion happens, Geminga could produce a much higher positron flux at Earth than the normal diffusion model, owing to the nature of L\'evy flight \citep{Wang:2021xph}. In this case, Geminga could significantly contribute to the positron excess even if the small diffusion coefficient measured around Geminga is applied in the whole region between Geminga and Earth.

The positron flux from Geminga is also determined by the positron injection spectrum. As introduced in Section~\ref{sec:injection}, the injection spectrum can be constrained by the spectral measurements of the pulsar halo. Note that the positron excess happens in the range of $\sim10-500$~GeV. Positrons in this energy range emit GeV photons through the IC scattering. Thus, only GeV observations of the Geminga halo could effectively constrain Geminga's contribution to the positron excess. \citet{Shao-Qiang:2018zla} analyzed the 10-yr Fermi-LAT data in the direction of Geminga and did not detect extended emission under different input parameters. The flux ULs seriously limit the positron injection power in the energy range of interest, which disfavors Geminga as the dominant source of the positron excess. On the other hand, \citet{DiMauro:2019yvh} also analyzed the Fermi-LAT data and claimed the detection of the GeV halo of Geminga. Compared with \citet{Shao-Qiang:2018zla}, they adopted a larger window for the analysis and took the proper motion of Geminga into account. Based on this measurement, \citet{DiMauro:2019yvh} claimed that Geminga alone still cannot interpret the positron excess due to the mismatch of the spectral shape. However, the calculation of \citet{2022arXiv220507038Z} indicates that Geminga is still likely to account for the positron excess under the constraint of the same gamma-ray observation. The difference may result from different parameter settings. As the Geminga halo is expected to be very extended in GeV, the spectral measurement is challenging owing to the large uncertainty of the gamma-ray background. More careful analyses are needed to cross-check the current measurements.

\subsection{Other Candidate Pulsars}
\label{subsec:candidates}
The latest positron spectrum measured by the AMS collaboration shows a sharp spectral dropoff around 300~GeV with a significance of more than $3\sigma$ \citep{Aguilar:2019owu}. This result provides important clues for the origin of the positron excess. The spectral cutoff can hardly be interpreted by the superposition of multiple pulsars \citep{Manconi:2020ipm}. Even old PWNe like the Geminga PWN can accelerate electrons to $\sim100$~TeV, so it is unreasonable to assume a typical cutoff energy of $\lesssim1$~TeV for the injection spectrum of PWNe to interpret the data. Instead, the spectral cutoff is more likely to correspond to the energy loss of the positrons released by a single source. In other words, the positron excess could be dominated by a single source \citep{Fang:2019ayz}.

Although Geminga is still a likely source of the positron excess, it is meaningful to investigate other possible dominant sources considering the slow-diffusion phenomenon. \citet{Fang:2019ayz} scanned all the identified pulsars\footnote{Millisecond pulsars are not included in the analysis of \citet{Fang:2019ayz}, while the nearest millisecond pulsar, PSR~J0437$-$4715, is proposed to be a likely source of the positron excess \citep{Bykov:2019mis}.} and searched for the pulsars that can both interpret the positron intensity and the spectral cutoff of the AMS-02 spectrum. The right panel of Figure~\ref{fig:positron} shows the contours of the positron flux at 300~GeV as a function of pulsar age and distance. Pulsars inside the shaded areas are those can contribute more than half of the measured flux of AMS-02 at 300~GeV. Three different models are presented with different colors: the one-zone slow-diffusion model, the one-zone fast diffusion model, and a two-zone diffusion model with $r_\star=50$~pc. The red band indicates the range of the energy-loss time corresponding to positron spectral cutoff. Thus, the candidate pulsars should lie in the intersection region between a shaded area and the red band. As can be seen, Geminga and PSR~B1055$-$52 are the most likely candidates. Another interesting finding is that the positron excess may not be interpreted by pulsars without the assumption of slow-diffusion zones, since the blue shaded area has no overlap with the red band.

PSR~B1055$-$52 is a Geminga-like pulsar, while its distance was thought to be larger than 1~kpc. The latest dispersion measure derives a close distance of 90 pc for PSR~B1055$-$52 \citep{2017ApJ...835...29Y}, and it becomes a very bright positron source. The annual parallax measurement for PSR~B1055$-$52 is suggested to obtain a more precise distance. The Fermi-LAT observation for extended emission is also essential to constrain its injection power.

\section{Diffuse TeV Gamma-ray Excess}
\label{sec:diffuse}
In this section, we briefly introduce the possible connection between pulsar halos and the diffuse gamma-ray excess. The main origin of the diffuse gamma-ray emission from the Galactic plane is generally believed to be the $\pi^0$-decay process from the interaction between the propagating CR nuclei and the ISM \citep{Strong:2007nh}. The diffuse gamma-ray emission below 100~GeV measured by Fermi-LAT and the local CR nuclei spectrum can be consistently interpreted under the homogeneous CR propagation model \citep{Fermi-LAT:2012edv}. At the TeV energy region, however, the $\pi^0$-decay emission predicted by the local CR spectrum is significantly lower than the diffuse gamma-ray measurements of Milagro and ARGO-YBJ experiments \citep{Milagro:2005xqq,Abdo:2008if,ARGO-YBJ:2015cpa}, which is known as the diffuse TeV gamma-ray excess \citep{Prodanovic:2006bq,2018PhRvL.120l1101L}. Spatially dependent propagation models are constructed to solve this contradiction \citep{Gaggero:2014xla}.

Alternatively, the excess could be attributed to unresolved gamma-ray sources. \citet{2018PhRvL.120l1101L} pointed out that pulsar halos and extended gamma-ray PWNe could be promising candidates for this unknown component. If all the young and middle-aged pulsars are associated with extended gamma-ray sources, an average conversion efficiency of $\sim10\%$ from the pulsar spin-down energy to the high-energy electrons could be enough to account for the excess \citep{2018PhRvL.120l1101L}. This model can naturally avoid the constraint from the local CR nuclei spectrum. Moreover, the gamma-ray spectra of pulsar halos and young PWNe are much harder than the diffuse $\pi^0$-decay spectrum, which means that this component may account for the TeV measurements while keeping consistency with the Fermi-LAT observations. On the other hand, the potential contribution of pulsar halos for the excess may vary with the mechanism of pulsar halos introduced in Section~\ref{sec:origin}. One may refer to \citet{2022arXiv220704011L} for a comprehensive discussion on this topic.

Recently, diffuse gamma-ray above 100~TeV was reported by the Tibet AS+MD experiment \citep{TibetASgamma:2021tpz}. If the $\pi^0$-decay component dominates the diffuse emission, there will be tension between the Tibet AS+MD measurement and the local CR spectrum even under the spatially dependent CR propagation model \citep{Qiao:2021iua}. \citet{Liu:2021lxk} pointed out that the upper limit of the Galactic neutrino emission may constrain the $\pi^0$-decay component, also indicating the existence of extra sources. The LHAASO experiment will test whether pulsar halos could contribute above 100~TeV in the near future. The sub-PeV excess may also be interpreted by the potential PeV accelerators, such as massive star clusters, hypernova remnants, or young PWNe \citep{Liu:2021lxk}.

\section{Future Works}
\label{sec:perspect}
Our understanding of pulsar halos is still in the preliminary stage. Future studies on pulsar halos should be carried out in both depth and breadth. Deeper observations of bright pulsar halos can provide insight into the CR propagation mechanism at small scales and then the properties of the MHD turbulence in the ISM. At the same time, a larger sample may also help us understand the origin of pulsar halos and the relation between local and global propagation of Galactic CRs.

\subsection{Deeper Studies on Bright Pulsar Halos}
\label{subsec:depth}
Morphology measurements of bright and nearby pulsar halos like the Geminga halo can provide the richest details about small-scale CR propagation. As shown in Figures~\ref{fig:profiles} and \ref{fig:prop}, the gamma-ray profiles predicted by different propagation models have different characteristics, while the current experimental results may not distinguish them. For the superdiffusion scenario, only the models with $\alpha$ close to $1$ are strongly disfavored. For the relativistic-modified fast-diffusion scenario, the goodness-of-fit test disfavors the model at a confidence level of $98.6\%$ for the Geminga halo case \citep{Bao:2021hey}, which has not reached the $3\sigma$ level. The ongoing LHAASO experiment \citep{Ma:2022aau} is expected to give more definite judgments to these models with better sensitivity and angular resolution. LHAASO may also be able to detect weak asymmetry in extensive halos, which is expected by the anisotropic diffusion model. Thus, more accurate morphology measurements are essential to test the mechanisms proposed in Sections~\ref{subsec:anisotropy} and \ref{subsec:relativistic} and hence determine whether pulsar halos originated from slow diffusion.

Energy-dependent morphology measurement of pulsar halos is not available at present but will be an essential subject for future study. This measurement determines the energy dependency of the CR diffusion coefficient, which reflects the properties of the MHD turbulence in the ISM and may also indicate the origin of pulsar halos. If the turbulence originated from the Alfv\'enic or slow magnetosonic cascade, the energy dependency could be $D(E)\propto E^{1/3}$ \citep{Goldreich:1994zz,Cho:2002qi}, while the fast magnetosonic scenario predicts $D(E)\propto E^{1/2}$ \citep{Cho:2002qi}. For the self-generated scenario, the energy dependency of $D$ is more complicated and may not be described by a power law \citep{Evoli:2018aza,Mukhopadhyay:2021dyh}. Current observations suggest that the injection spectra of pulsar halos have a cutoff at $\sim100$~TeV. As the turbulence growth rate is positively correlated with the injection power under the self-generated mechanism, a sharp increase of $D$ is predicted at high energies \citep{Mukhopadhyay:2021dyh}. Besides, $D(E)\propto E$ may correspond to the Bohm diffusion, which means that the turbulent growth has reached saturation.

The energy spectra of pulsar halos are measured in a certain angular range and thus determined by both the injection spectra and $D(E)$. Energy-dependent morphology measurements are indispensable to decouple the injection spectra and $D(E)$. LHAASO-KM2A combined with the Water Cherenkov Detector Array of LHAASO (LHAASO-WCDA) can measure the morphologies of pulsar halos in the energy range from $\sim1$~TeV to $\sim1$~PeV, which could be wide enough to understand the forms of the injection spectra and $D(E)$. Bright pulsar halos are feasible targets for energy-dependent measurements as it is possible to divide more energy bins for these sources.

We should note that the comparison between the localized diffusion coefficient inferred from pulsar halos and the global diffusion coefficient inferred from B/C is based on the energy extrapolation of the result measured by B/C. For example, the global diffusion coefficient suggested by the diffusion-reacceleration model of \citet{Yuan:2017ozr} is $\approx700$ times larger than the value in the Geminga halo if extrapolated to 100~TeV. However, the B/C measurement is only up to $\sim1$~TeV at present, and the energy dependency of the global diffusion coefficient is not available at higher energies. On the one hand, the ongoing DArk Matter Particle Explorer \citep[DAMPE]{DAMPE:2017cev} will provide multi-TeV measurement of B/C in the coming future. We may also expect future space experiments like the High Energy cosmic-Radiation Detection \citep[HERD]{HERD:2014bpk} facility to boost the B/C measurement to $\sim100$~TeV. On the other hand, we may measure the sub-TeV diffusion coefficient in pulsar halos by Fermi-LAT or H.E.S.S to give a straightforward comparison with the Galactic-averaged value.

\subsection{A Larger Sample of Pulsar Halos}
\label{subsec:breadth}
When a large sample of pulsar halos is available, we can statistically investigate the correlations between the parameters of pulsar halos. This is helpful in judging the origin of pulsar halos. The self-generated model will be disfavored if there is no correlation between the diffusion coefficient and the electron injection power. For example, as noticed by \citet{2022arXiv220713533F}, the diffusion coefficient in HESS~J1831$-$098 is comparable to the other known pulsar halos, while the spin-down luminosity of PSR~J1831$-$0952 is dozens of times larger than the other pulsars. This implies that the slow-diffusion zone around PSR~J1831$-$0952 may not be self-generated. Another possibility is that the slow-diffusion zone \textit{is} self-generated while the turbulent growth has reached saturation for the known pulsar halos. The latter case will be possible if $D(E)\propto E$. 

At the same time, if the diffusion coefficient in pulsar halos is independent of the position in the Galaxy, the slow-diffusion environment is more likely to be self-generated or produced by the associated SNR rather than the superposition effect of multiple external turbulent sources. An example is LHAASO~J0621$-$3755. It is far from the Galactic plane, while its diffusion coefficient is comparable to the other pulsar halos near the Galactic plane. It indicates that the slow diffusion may not only occur in the Galactic disk but as a localized phenomenon around some middle-aged pulsars. Moreover, searching for extended gamma-ray emission around millisecond pulsars is important to test the SNR-generated scenario as these pulsars have been far away from their parent SNRs.

The LHAASO experiment is expected to discover dimmer pulsar halos owing to the better sensitivity. This is crucial for inferring the statistical properties of pulsar halos. Assuming a fixed electron injection power, pulsar halos with a larger diffusion coefficient will be dimmer. Thus, if the sample only includes the brightest pulsar halos, the inference will be seriously affected by the selection effect. 

In addition, the local environment of pulsars, such as the interstellar radiation field and the magnetic field strength, can also affect the brightness of pulsar halos. These factors should be considered in the statistical analyses. If the magnetic field strength has a significant spatial variation in the local environment, the electron spatial distribution will be affected, hence the diffusion coefficient measurement. Observations of diffuse synchrotron emission in the X-ray band may help to test this scenario \citep{Li:2021nrm}.

\subsubsection{Are Pulsar Halos Universal?}
\label{subsubsec:universal}
The universality of pulsar halos is often mentioned in the prospect of future observations \citep{Linden:2017vvb,Sudoh:2019lav}. It may answer whether the localized slow-diffusion around pulsars is related to the inferred spatially dependent CR diffusion at the Galactic scale as discussed in Section~\ref{subsec:prop_global} and how pulsar halos contribute to the excess of the diffuse TeV emission introduced in Section~\ref{sec:diffuse}.

Current observations indicate that pulsar halos may not be universal. PSR~J1809$-$2332, B0906$-$49, and J1105$-$6107 are the brightest middle-aged pulsars within the survey region of the H.E.S.S. Galactic plane survey (HGPS). Their $L/D^2$ is more than $5$ times larger than that of PSR~J1831$-$0952, where $L$ and $D$ are the pulsar spin-down luminosity and distance, respectively. Significant TeV emission is expected around these pulsars if pulsar halos are universal; however, no signal is detected by HGPS \citep{HESS:2018pbp}. Moreover, if all the nearby middle-aged pulsars have the same slow-diffusion environment and injection efficiency as the Geminga halo, the predicted positron flux at Earth will be much higher than observed \citep{Fang:2019ayz,2022arXiv220611803M}. Based on this, \citet{2022arXiv220611803M} point out that the occurrence rate of pulsar halos could be $\sim5-10\%$.

It should be noted that the non-universality of pulsar halos is not equal to the non-universality of the slow-diffusion environment around middle-aged pulsars. The brightness of a pulsar halo is also determined by the electron acceleration efficiency and escape efficiency of the center PWN. The reason for the invisibility of pulsar halos is not necessarily the same for different objects. For example, among the three bright pulsars mentioned in the last paragraph, PSR~J1809$-$2332 has a significant bow-shock X-ray PWN \citep{VanEtten:2012jg}. It indicates that this source can indeed generate high-energy electrons, and the invisibility of the pulsar halo is possibly due to the low escape efficiency from the PWN or the lack of slow-diffusion environment. However, no X-ray PWN is observed for the other two pulsars, which means that their ability to generate high-energy electrons is questionable.

We should also note that although the factors of the slow diffusion and acceleration efficiency may be degenerate for the visibility of a pulsar halo, the implication could be very different for the interpretation of positron excess or the diffuse gamma-ray emission. In the absence of slow diffusion, the electrons/positrons generated by the pulsar may still contribute to the positron excess or the diffuse emission, while the opposite is true in the absence of high-energy electron acceleration.

\section{Summary}
\label{sec:sum}
We briefly summarize the key points of this review as follows:

\begin{enumerate}
    \item Pulsar halos are a new class of gamma-ray sources represented by the Geminga halo. They are generated by electrons diffusing in the ISM around pulsars. It is essential to distinguish them from PWNe: PWNe should be regarded as the electron sources of pulsar halos.
    \item The diffusion inferred from the morphologies of pulsar halos is extremely slow, while the origin of the slow-diffusion environment is still uncertain. It may be self-generated or left by external sources like SNRs. Some other models may interpret the pulsar halos without slow diffusion.
    \item Pulsar halos are 'microscopes' for studying the Galactic CR propagation, which can test sophisticated propagation models at small scales. They are complementary to the global probes of Galactic CR propagation like B/C.
    \item Pulsar halos are essential for inferring the electron injection spectra from PWNe, which help us understand the particle escape process from PWNe.
    \item The discovery of pulsar halos has a significant influence on the interpretation of the positron excess. Geminga is still the most likely source of the positron excess. Deeper studies of the Geminga halo at $\sim10$~GeV are crucial to this issue.
    \item Pulsar halos may contribute to the unknown excess of the diffuse TeV gamma-ray emission.
    \item Pulsar halos are probably not universal, which may be attributed to the non-universality of the slow-diffusion environment or the difference in the electron injection efficiency of PWNe.
\end{enumerate}

\section*{Conflict of Interest Statement}

The authors declare that the research was conducted in the absence of any commercial or financial relationships that could be construed as a potential conflict of interest.


\section*{Funding}
This work is supported by the National Natural Science Foundation of China under Grants No. 12105292, and No. U1738209.

\section*{Acknowledgments}
The author thanks Prof. Xiao-Jun Bi for providing motivation for writing this paper. The author also thanks the reviewers for their detailed and valuable suggestions.

\bibliographystyle{Frontiers-Harvard} 
\bibliography{references}


\end{document}